\documentclass[runningheads]{llncs}
\usepackage[T1]{fontenc}
\usepackage{graphicx}
\usepackage{syntax}
\usepackage{graphicx}
\usepackage{booktabs}
\usepackage{subcaption}
\usepackage{mathtools}
\usepackage{dsfont}
\usepackage{paralist}
\usepackage{caption}
\usepackage{amscd}
\usepackage{tikz}
\usetikzlibrary{matrix}
\usetikzlibrary{fit}
\usetikzlibrary{automata}
\usetikzlibrary{arrows,shapes,backgrounds,positioning,calc,decorations.pathmorphing} 
\usepackage{multirow}
\usepackage{dashbox}
\usepackage{amsfonts}
\usepackage[normalem]{ulem}
\usepackage{wrapfig}
\usepackage{appendix}
\usepackage{algpseudocode}
\usepackage[linesnumbered,ruled,noend]{algorithm2e}
\algrenewcommand\textproc{\texttt} % to make function calls not in small caps
\usepackage{amsmath,amssymb} %for the \xrightsquigarrow
\algnewcommand\algorithmicforeach{\textbf{for each}}
\algdef{S}[FOR]{ForEach}[1]{\algorithmicforeach\ #1\ \algorithmicdo}
\usepackage{listings}
\usepackage[utf8]{inputenc}
\usepackage[T1]{fontenc}
\usepackage{textcase}
\usepackage{empheq}
\usepackage{tcolorbox}
\usepackage{stmaryrd}

\usepackage{xspace}

\newcommand{\trans}[3]{\ensuremath{#1 \xrightarrow{#2} #3}\xspace}

\newcommand{\eg}{e.g.\xspace}
\newcommand{\ie}{i.e.\xspace}

\newcommand{\LS}{S}
\newcommand{\ls}{s}

\newcommand{\freech}{independent\xspace}

\makeatletter
\newsavebox{\@brx}
\newcommand{\llangle}[1][]{\savebox{\@brx}{\(\m@th{#1\langle}\)}%
  \mathopen{\copy\@brx\kern-0.5\wd\@brx\usebox{\@brx}}}
\newcommand{\rrangle}[1][]{\savebox{\@brx}{\(\m@th{#1\rangle}\)}%
  \mathclose{\copy\@brx\kern-0.5\wd\@brx\usebox{\@brx}}}
\makeatother

\newcommand{\lemref}[1]{Lemma~\ref{lem:#1}}

\newcommand{\secref}[1]{Sec.~\ref{sec:#1}}
\newcommand{\figref}[1]{Fig.~\ref{fig:#1}}

\newcommand{\algoref}[1]{Algo.~\ref{algo:#1}}

\newcommand{\appref}[1]{App.~\ref{app:#1}}

\newcommand{\lineref}[1]{Line~\ref{line:#1}}

\newcommand{\para}[1]{\noindent{\bf \em #1}}

\newcommand{\powerset}{\mathcal{P}}

\newcommand{\LQ}{S}
\newcommand{\GQ}{Q}
\newcommand{\iGQ}{Q_\interpretation}
\newcommand{\loq}{s}
\newcommand{\gq}{q}

\newcommand{\T}{T}

\newcommand{\GT}{R}
\newcommand{\iGT}{R_\interpretation}
\newcommand{\iDisGT}{R_\interpretation^{\text{\textit{dis}}}}

\newcommand{\lhs}{\mathtt{lhs}}
\newcommand{\rhs}{\mathtt{rhs}}

\newcommand{\nats}{\mathbb{N}}

\newcommand{\sysname}{QuickSilver}
\newcommand{\kinarach}{\textsc{\sysname}\xspace}

\newcommand{\permissible}{permissible\xspace} 
 
\newcommand{\amenable}{cutoff-amenable\xspace}

\newcommand{\pairwise}{rendezvous\xspace}

\definecolor{Gray}{gray}{0.85}
\definecolor{LightRed}{rgb}{1,0.7,0.7}
\definecolor{LightGreen}{rgb}{0.7,1,0.7}

\definecolor{realGreen}{rgb}{0.0, 0.5, 0.0}

\newcommand{\mn}[1]{}

\newcommand{\mnx}[1]{}

\newcommand{\aml}{\textsc{Mercury}\xspace}

\newcommand{\chwellbehaved}{phase-compatible\xspace} 
\newcommand{\chwellbehavedness}{phase-compatibility\xspace} 
 
\newcommand{\CHwellbehavedness}{Phase-Compatibility\xspace}

\newcommand{\amenability}{cutoff-amenability\xspace} 
\newcommand{\Amenability}{Cutoff-amenability\xspace}
\newcommand{\AMenability}{Cutoff-Amenability\xspace}

\newcounter{sarrow}
\newcommand\xrightsquigarrow[1]{%
\stepcounter{sarrow}%
\mathrel{\begin{tikzpicture}[baseline= {( $ (current bounding box.south) + (0,-0.5ex) $ )}]
\node[inner sep=.5ex] (\thesarrow) {$\scriptstyle #1$};
\path[draw,<-,decorate,
  decoration={zigzag,amplitude=0.7pt,segment length=1.2mm,pre=lineto,pre length=4pt}] 
    (\thesarrow.south east) -- (\thesarrow.south west);
\end{tikzpicture}}%
}

\newcommand{\locVar}{v_{loc}}
\newcommand{\valOf}[2]{#1(#2)}

\newcommand{\termb}[1]{\code{#1}}
\newcommand{\terma}[1]{\textbf{\code{#1}}}

\newcommand{\ourskip}{\smallskip}

\newcommand{\actb}{\text{\termb{f}}}

\newcommand{\gsactions}{E_\mathtt{{global}}}

\newcommand{\pha}{ph}

\newcommand{\states}{st}

\newcommand{\firable}{initiable\xspace}

\newcommand{\uf}{\code{uf}}
\newcommand{\figHole}[1]{{\color{red}\terma{??}$_\text{\terma{#1}}$}}
\newcommand{\ufDSL}[1]{\terma{??}_\text{\terma{#1}}}

\newcommand{\process}{P}
\newcommand{\psketch}{\process_{\text{\textit{sk}}}}
\newcommand{\pcomplete}{\process_{\interpretation}}
\newcommand{\pcompletee}[1]{\process_{\interpretation,#1}}

\newcommand{\semantics}[1]{\llbracket#1\rrbracket}
\newcommand{\globalSemantics}[2]{\llbracket#1,#2\rrbracket}

\newcommand{\isemantics}[1]{\semantics{#1}}
\newcommand{\iglobalSemantics}[1]{\globalSemantics{#1}}

\newcommand{\safetySpec}{\phi_s(n)}
\newcommand{\safetySpecc}{\phi_s(c)}
\newcommand{\livenessSpec}{\phi_l(c)}
\newcommand{\phaseCompSpec}{\phi_{pc}}
\newcommand{\cutoffsSpec}{\phi_{ca}}

\newcommand{\mpsp}{MPSP\xspace}

\newcommand{\interStartSymb}{\llangle}
\newcommand{\interEndSymb}{\rrangle}

\newcommand{\existsTr}[1]{\interStartSymb #1 \interEndSymb_\interpretation}
\newcommand{\reachable}[1]{\interStartSymb #1 \interEndSymb_\interpretation}
\newcommand{\hasAction}[2]{\interStartSymb #2:#1 \interEndSymb_\interpretation}
\newcommand{\hasNoAction}[2]{\interStartSymb \neg #2:#1 \interEndSymb_\interpretation}
\newcommand{\goesTo}[3]{\interStartSymb #3:#1,#2 \interEndSymb_\interpretation}

\newcommand{\srcSet}[1]{src_#1}
\newcommand{\destSet}[1]{dst_#1}
\newcommand{\firableIn}[2]{\firable(#1,#2)}

\newcommand{\samePhase}[3]{inPhase(#1,#2,#3)}

\newcommand{\act}{\text{\termb{e}}}

\newcommand{\prop}{\phi}

\newcommand{\cex}{cex}

\newcommand{\learner}{learner\xspace}

\newcommand{\teacher}{teacher\xspace}

\newcommand{\vars}{V}
\newcommand{\var}{v}
\newcommand{\event}{e}
\newcommand{\events}{E}

\newcommand{\actingPolarity}{A}
\newcommand{\reactingPolarity}{R}
\newcommand{\acting}[1]{\actingPolarity(#1)}
\newcommand{\reacting}[1]{\reactingPolarity(#1)}
\newcommand{\handlers}{H}
\newcommand{\handler}{h}

\newcommand{\guard}{g}
\newcommand{\update}{u}
\newcommand{\action}{a}
\newcommand{\handlerAction}{\action}

\newcommand{\uFuns}{F_{\text{\textit{sk}}}}
\newcommand{\genericUFun}{f}
\newcommand{\uFun}{\genericUFun_{\text{\textit{sk}}}}
\newcommand{\interpretation}{I} %\newcommand{\interpretation}{I_{\text{\textit{sk}}}} 
\newcommand{\interpretations}{\mathcal{I}_{\text{\textit{sk}}}} 

\newcommand{\allConstraints}{\mathcal{C}} 
\newcommand{\constraints}{C}
\newcommand{\constraint}{\mathtt{c}}

\newcommand{\argument}{\mathtt{arg}}

\newcommand{\itransitions}{\T_\interpretation}
\newcommand{\idistransitions}{\itransitions^{\text{\textit{dis}}}}
\newcommand{\idistransitionsPrime}{\itransitions'^{\text{\textit{dis}}}}
\newcommand{\istates}{\LQ_\interpretation}

\newcommand{\cexPhComp}{\text{\textit{cex}}_p} 
 
\newcommand{\cexSafety}{\text{\textit{cex}}_s}

\newcommand{\activeTrs}{active}

\newcommand{\tool}{\textsc{Cinnabar}\xspace}

\newcommand{\cube}{c}
\newcommand{\literal}{l}
\newcommand{\literalWitness}{lw}
\newcommand{\cubeWitness}{cw}

\newcommand{\buchi}{B\"{u}chi\xspace}

\newcommand{\fair}{fair\xspace}
\newcommand{\Fair}{Fair\xspace}

\newcommand{\effdecidable}{efficiently-decidable\xspace}

\newcommand{\EFfdecidable}{Efficiently-Decidable\xspace}

\newcommand{\counterexample}{counterexample\xspace}
\newcommand{\Counterexample}{Counterexample\xspace}
\newcommand{\counterexamples}{counterexamples\xspace}
\newcommand{\Counterexamples}{Counterexamples\xspace}

\newlength{\myMheight}
\definecolor{bluekeywords}{rgb}{0.13,0.13,1}
\definecolor{greencomments}{rgb}{0,0.5,0}
\definecolor{turqusnumbers}{rgb}{0.17,0.57,0.69}
\definecolor{redstrings}{rgb}{0.5,0,0}

\definecolor{qualifiers}{rgb}{0.5,0,0.63}
\definecolor{types}{rgb}{0.5,0,0}
\definecolor{communication}{rgb}{0.5,0,0}
\definecolor{categories}{rgb}{0.5,0,0.63}
\definecolor{controlstmts}{rgb}{0.5,0,0}
\definecolor{handlermeta}{rgb}{0.13,0.13,1}
\definecolor{setops}{rgb}{0.5,0,0}
\definecolor{cons}{rgb}{0.5,0,0}
\definecolor{hlcolor}{rgb}{0.13,0.13,1}

\lstdefinelanguage{FSharp}{
	keywords={},
	otherkeywords={
	??,
	??<1>
	},
    keywordstyle=[1]\bfseries\color{red},
    keywords=[2]{env, initial}, 
    keywordstyle=[2]\bfseries\color{qualifiers},
    keywords=[3]{int, idset, id, bool, unit},
    keywordstyle=[3]\bfseries\color{types}, % this can be used to have other words show in a different style.
    keywords=[4]{sendrz, sendbr,recv,br,rz}, 
    keywordstyle=[4]\bfseries\color{communication},
    keywords=[5]{machine, variables, actions,events, location, satisfies,process},
    keywordstyle=[5]\bfseries\color{categories},
    keywords=[6]{if,else,goto},
    keywordstyle=[6]\bfseries\color{controlstmts},
    keywords=[7]{on, win, lose, passive, do, where},
    keywordstyle=[7]\bfseries\color{handlermeta},
    keywords=[8]{add, remove, con=tains,decVar,winS,loseS,payld, default},
    keywordstyle=[8]\bfseries\color{setops},
    keywords=[9]{partition, consensus},
    keywordstyle=[9]\bfseries\color{cons},
    sensitive=false,
    morecomment=[l][\color{greencomments}]{///},
    morecomment=[l][\color{greencomments}]{//},
    morestring=[b]",
    stringstyle=\color{redstrings}
    }    
    
    \lstnewenvironment{dsl}
  {
    \lstset{
        language=FSharp,
        basicstyle=\ttfamily,
         numbers=left,
    	stepnumber=1,
        breaklines=true,
        escapeinside={{(*}{*)}},
        columns=fullflexible
        }
        }
  {
  }

    \lstnewenvironment{udsl}
  {
    \lstset{
        language=FSharp,
        basicstyle=\ttfamily,
        breaklines=true,
        escapeinside={{(*}{*)}},
        columns=fullflexible
        }
        }
  {
  }

    \lstnewenvironment{sdsl}
  {
    \lstset{
        language=FSharp,
        basicstyle=\ttfamily\footnotesize,
         numbers=left,
    	stepnumber=1,
        breaklines=true,
        escapeinside={{(*}{*)}},
        columns=fullflexible
        }
        }
  {
  }

      \lstnewenvironment{ssdsl}
  {
    \lstset{
        language=FSharp,
	 basicstyle=\ttfamily\scriptsize,
         numbers=left,
    	stepnumber=1,
        breaklines=true,
        escapeinside={{(*}{*)}},
        columns=fullflexible
        }
        }
  {
  }

     \lstnewenvironment{sudsl}
  {
    \lstset{
        language=FSharp,
        basicstyle=\ttfamily\scriptsize,
        breaklines=true,
        escapeinside={{(*}{*)}},
        columns=fullflexible
        }
        }
  {
  }

\newcommand{\code}[1]{\texttt{#1}} %at least for now
\newcommand{\reducespace}{-1pt}
\newcommand{\squeezecaption}{\vspace{\reducespace}}

\usepackage{hyperref}
\begin{document}
\title{Synthesis of Distributed Agreement-Based Systems with Efficiently-Decidable 
%Parameterized 
Verification}
%\title{}
%\title{Contribution Title\thanks{Supported by organization x.}}
%
\titlerunning{\tool}
%\author{First Author\inst{1}\orcidID{0000-1111-2222-3333} \and
%Second Author\inst{2,3}\orcidID{1111-2222-3333-4444} \and
%Third Author\inst{3}\orcidID{2222--3333-4444-5555}}
%% First names are abbreviated in the running head.
%% If there are more than two authors, 'et al.' is used.
%%
%\institute{Princeton University, Princeton NJ 08544, USA \and
%Springer Heidelberg, Tiergartenstr. 17, 69121 Heidelberg, Germany
\author{
	Nouraldin Jaber\inst{1} \and
	Christopher Wagner\inst{1} \and
	Swen Jacobs\inst{2} \and
	Milind Kulkarni\inst{1} \and
	Roopsha Samanta\inst{1} 
}

\authorrunning{Jaber et al.}
%% First names are abbreviated in the running head.
%% If there are more than two authors, 'et al.' is used.
%%
\institute{Purdue University, West Lafayette, USA \\\email{\{njaber,wagne279,milind,roopsha\}@purdue.edu} \and
	CISPA Helmholtz Center for Information Security, Saarbr\"{u}cken, Germany
	%CISPA, Saarbr\"{u}cken, Germany
	\email{jacobs@cispa.de}
}
%\institute{}
%
%
\maketitle              % typeset the header of the contribution
%
%Alternative 2-line titles
%\begin{compactenum}
%\item Synthesis of Distributed Agreement-Based Systems with Efficiently-Decidable Verification.
%\item Synthesizing Correct Distributed Agreement-Based Systems.
%\item Synthesizing Distributed Agreement-Based Systems with Decidable Verification.
%\end{compactenum}
\begin{abstract}
%\mn{Removed ''Parameterized'' from title.}
{\em Distributed agreement-based} (DAB) systems use common distributed agreement protocols such as leader election and consensus as building blocks for their target functionality. While automated verification for DAB systems is undecidable in general, recent work identifies a large class of DAB systems for which verification is {\em efficiently-decidable}.
Unfortunately, the conditions characterizing such a class can be opaque and non-intuitive, and can pose a significant challenge to system designers trying to model their systems in this class.

In this paper, we present a synthesis-driven tool, \tool, 
%to help system designers building DAB systems ``fit'' their intended designs into an efficiently-decidable class.
to help system designers building DAB systems ensure that their intended designs belong to an efficiently-decidable class.
In particular, starting from an initial \emph{sketch} provided by the designer, \tool generates sketch completions using a counterexample-guided procedure. The core technique relies on compactly encoding root-causes of counterexamples to varied properties such as efficient-decidability and safety. 
We demonstrate \tool's effectiveness by successfully and efficiently synthesizing completions 
for a variety of interesting DAB systems including a distributed key-value store and a distributed consortium system.
%\keywords{First keyword  \and Second keyword \and Another keyword.}
\end{abstract}
\section{Introduction}
\label{sec:intro}

Distributed system designers are increasingly embracing the incorporation of formal verification techniques into their development pipelines~\cite{aws1,aws2,moveProver,googleFolks}.
The formal methods community has been enthusiastically responding to this trend with a wide array of modeling and verification frameworks for prevalent distributed systems~\cite{Padon.IvySafetyVerification.PLDI.2016,Jaber.QuickSilver.OOSLA.2021,Hawblitzel.Ironfleet.SOSP.2015,Sergey.ProgrammingProvingDistributed.POPL.2017}.
%
%supporting verification strategies that mirror the modular, parameterized nature of distributed systems \cite{Padon.IvySafetyVerification.PLDI.2016, Jaber.QuickSilver.OOSLA.2021,Hawblitzel.Ironfleet.SOSP.2015,Sergey.ProgrammingProvingDistributed.POPL.2017}.
%
 %distributed services are built on top of common, core distributed protocols (e.g., leader election, consensus, two-phase commit) many of these frameworks enable {\em modular} design and verification
%arately reason about the correctness of the core protocols, then replace them with simpler abstractions when reasoning about overall system.
% 
%Ideally,
A desirable workflow for a system designer using one of these frameworks is to (1) provide a framework-specific model and specification of their system, and (2) \emph{automatically} verify if the system model meets its specification.
%Ideally, the hope is that all the system designer needs to do is to pick their
%favorite verification framework, model their system in the modeling language
%provided by that framework, provide correctness properties of their system,
%and for the verification framework to check if the model meets its
%specifications. 
%

However, the problem of algorithmically checking if a distributed system is correct for an {\em arbitrary} number of processes, \ie, the {\em automated parameterized verification problem}, is undecidable, even for finite-state processes~\cite{kozan.pmcp.undecidable.1986,Suzuki.PMCP.UndecidableFirstPPR.1988.PPR}. To circumvent undecidability, the system designer must be involved, {\em one way or another}, in the verification process.
{\em Either} the designer may choose a semi-automated verification
approach and use their expertise to ``assist'' the verifier by providing inductive invariants \cite{Sergey.ProgrammingProvingDistributed.POPL.2017,dafny,Hawblitzel.Ironfleet.SOSP.2015,Wilox.Verdi.PLDI.2015}.
{\em Or,} the designer may choose a fully-automated verification approach that is only applicable to a restricted class of system models~\cite{GSP,Jaber.QuickSilver.OOSLA.2021,Lazic.SynthesisDistributedAlgorithms.X.2018,BloemETAL15} and use their expertise to 
%``fit'' the model of their system into the decidable class. 
ensure that the model of their system belongs to the decidable class. 
This begs the question---{\em for each workflow, how can we further simplify the system designer's task?}
While effective frameworks have been developed to aid the designer in discovering inductive invariants for the first workflow (\eg, Ivy~\cite{Padon.IvySafetyVerification.PLDI.2016}, I4~\cite{I4}), there has been little emphasis on aiding the designer to build {\em decidability-compliant} models of their systems for the second workflow. 
% aware
% yielding
% acquiescent
% compliant
% accommodating

%so that the resulting system model and invariants fall within a decidable fragment of first-order logic, rendering the verification problem decidable. 

%Providing \neww{and/or manipulating} inductive invariants, however, remains a challenging task for a system designer who may not have the necessary background to formally reason about their systems.
%
%
%While the designer is not required to provide an inductive invariant to utilize these techniques, the hidden complexity lies in \emph{fitting} their system model into the class of models that supports decidable verification.

%\kinarach \cite{Jaber.QuickSilver.OOSLA.2021}, a framework for modeling and automated parameterized verification of \emph{distributed agreement-based} systems adopts such an approach. 
%
%\edit{In the first para, we want to introduce the general idea, not directly cinnabar. We want to do this in a way that does (i) does not make it sound like a huge contribution, and (ii) distinguish it from traditional synthesis (by saying we also fit it to be decidable). We want this strategy to look domain agnostic. The second para should be an instantiation that target the QS tool.}

In this paper, we present a synthesis-driven approach to help system designers using the second workflow to build models that are {\em both} decidability-compliant and correct. 
%Thus,  our approach helps designers to construct models that ``fit'' into a decidable class for automated,
Thus,  our approach helps designers to construct models that belong to a decidable class for automated, parameterized verification, and can be automatically verified to be safe for any number of processes. 
%in addition to helping designers build correct\mn{remove the word ``correct'' here because it might make people wonder ``why do I care about the rest?''} models \neww{(like traditional synthesis approaches)},

In particular, we instantiate this approach in a tool, \tool, that targets an existing framework, \kinarach, for modeling and automated verification of {\em distributed agreement-based (DAB) systems}~\cite{Jaber.QuickSilver.OOSLA.2021}. Such systems use {\em agreement protocols} such as leader election and consensus as building blocks. \kinarach enables modular verification of DAB systems by providing a modeling language, \aml, that allows designers to model {\em verified} agreement  protocols using inbuilt language primitives, and identifying a class of \aml models for which the parameterized verification problem is %not only decidable, but also 
\emph{efficiently decidable}.

%\begin{newtext}
%In this paper, we present a synthesis-driven tool, \tool, to help system designers using the second workflow to build models that are {\em both} decidability-compliant and correct. Thus, in addition to helping designers build correct models, \tool first helps designers ``fit'' their models into a decidable class for automated (parameterized) verification.
%
%In particular, \tool targets an existing framework, \kinarach, for modeling and automated verification of {\em distributed agreement-based (DAB) systems}~\cite{Jaber.QuickSilver.OOSLA.2021}. Such systems use {\em agreement protocols} such as leader election and consensus as building blocks. \kinarach enables modular verification of DAB systems by providing a modeling language, \aml, that allows designers to model {\em verified} agreement  protocols using inbuilt language primitives, and identifying a class of \aml models for which the parameterized verification problem is not only decidable, but also efficient. 
%\end{newtext}
%When a \aml model is in the \effdecidable fragment, \kinarach computes a \emph{cutoff} number of processes such that the parameterized verification of the modeled system with any number number of processes is reduced to the verification of a system with that cutoff number of processes.
%
%In particular, a pair $\langle \p,\phi\rangle$ of a \aml model $\p$ and safety specification $\phi$ is in the \effdecidable fragment of \aml if it satisfies 
%\neww{a sufficient set of 
%``\chwellbehavedness'' and ``\amenability'' 
%conditions.} 

Unfortunately, this \emph{\effdecidable} class of \aml models is characterized using conditions 
  %are carefully constructed to pave the way for decidability but are not necessarily \emph{user friendly}.
%a set of \emph{conditions}. 
that are rather opaque and non-intuitive, and can pose a significant challenge to system designers trying to model their systems in this class. 
The designer is responsible for understanding the conditions, and manually modifying their system model to ensure it belongs to the \effdecidable class of \aml. This process can be both tedious and error-prone, even for experienced system designers. %

%\tool targets designers of \aml models of DAB systems \neww{and demonstrates that synthesis can be used to build \aml models that belong to the \effdecidable fragment and are correct.}
\tool demonstrates that synthesis can be used to automatically build models of DAB systems that belong to the \effdecidable fragment of \aml and are correct.
%\nour{move this to description of procedure}
%Starting from an initial \emph{sketch} of the system design provided by the designer, \tool leverages synthesis techniques to generate \emph{sketch completions} that belong to the \effdecidable class of \aml \emph{and} are correct.\ourskip 
%

%In a nutshell, we propose a multi-stage, counter-example-guided inductive synthesis (CEGIS)-based architecture capable of completing a \aml sketch so that the resulting \aml model belongs to the \effdecidable fragment of \aml.\milind{I like how this summarization reads, but it feels pretty redundant with the previous paragraph.}\ourskip

\para{Contributions.} The key contributions of this paper are:

\begin{compactenum}

 %\nour{talk about the goal not the procedure} \emph{A counterexample-guided synthesis procedure for \aml models "a synthesis-driven method to ..."} (\secref{algorithm}). 
  
   \item \emph{A synthesis-driven method for building \effdecidable, correct \aml models} (\secref{algorithm}). \mn{This point is more of the "expected use of said method (i.e., its interface), as opposed to a design detail.} Starting from an initial \emph{sketch} of the system design provided by the designer, \tool generates a \emph{sketch completion} that (i) belongs to the \effdecidable class of \aml \emph{and}  (ii) is correct.
%(ii) if so, in the latter stages, computes a cutoff $c$ and checks if the system is safe 
 %(and live) 
% for that number $c$ of processes.
% If a candidate completion fails at any stage, the \learner proposes a new candidate to try.
 
 \item 
 %\mn{too wordy, can shorten depending on section 2} 
 \emph{A \counterexample-guided synthesis procedure that leverages an efficient, extensible, multi-stage architecture} (\secref{cinnabar}). We present a procedure that involves a \learner that proposes completions of the \aml sketch, and a \teacher that checks if the completed model belongs to the \effdecidable class of \aml and is correct.
 To enable efficient synthesis using this procedure, we propose an architecture that proceeds in stages. The initial stages focus on checking if a completed model is in the \effdecidable class while the latter stages focus on checking if a completed model is also correct. 
 To enable efficiency, when a candidate completion fails at any stage, the architecture helps the learner avoid `` similar'' completions by 
\emph{extracting} a \emph{root-cause} of the failure and \emph{encoding} the root-cause as an additional constraint for the learner.
Each stage is equipped with a \counterexample extraction strategy tailored to the {\em property} checked in that stage. 
The encoding procedure,
%\editt{find a way to relate it ti constraints. might just say ``the encoding procedure''}
on the other hand, is property-agnostic---it is able to encode the root-cause of any failure regardless of the stage that extracts it. The separation of the \counterexample extractions and the encoding allows the architecture to be extensible---one can add a new stage with a new \counterexample extraction strategy, and leverage the existing encoding.
    
 \item \emph{The \tool tool} (\secref{evalAndImpl}). We develop a tool, \tool, to help system designers build \aml models of DAB systems. 
 \tool employs \kinarach as its \teacher and the Z3 SMT solver as its \learner.
  \tool is able to successfully and efficiently complete \aml sketches of various interesting distributed agreement-based systems.\ourskip

\end{compactenum}

%\mn{revise if needed}\para{Organization.} We give an illustrative overview of our contributions in \secref{overview}. The synthesis problem for \aml systems in formalized in \secref{model} and our solution approach to it is found in \secref{algorithm}. Our tool, \tool, is presented in \secref{cinnabar} and its empirical evaluation is detailed in \secref{evaluation}.

%\input{overview}
\section{The \aml Parameterized Synthesis Problem}
\label{sec:model}

%\aml is a modeling language for distributed systems that build on top of verified agreement protocols such as leader election and consensus. 
%\aml provides two agreement primitives that abstract away the intricacies of agreement protocols and allows the system designer to focus on reasoning about the agreement-based system that uses these primitives. 

%In this section, 
%We now 
We first briefly review the syntax and semantics of \aml~\cite{Jaber.QuickSilver.OOSLA.2021}, 
a modeling language for distributed systems that build on top of verified agreement protocols such as leader election and consensus. 
Then, 
%we introduce the necessary extensions to the syntax and semantics that enable synthesis of \aml systems, and 
we formalize the synthesis problem.

\subsection{Review: \aml Systems}
\label{sec:mercury}
\para{\aml Process Definition.} 
A \aml 

\begin{wrapfigure}[11]{r}{0.37\linewidth}
\centering
\vspace{-4em}
\begin{tcolorbox}[colback=white,sharp corners,boxrule=0.3mm,top=-1mm,bottom=-1mm, left = 2mm, right=-1mm]
%, width=0.35\linewidth]
  %,top=-1.5mm,bottom=-1.5mm,right=0mm]
%\begin{tcolorbox}
\begin{sudsl}
process DistributedStore
variables 
  int[1,5] cmd
events
  env rz doCmd  : int[1,5]
initial location Candidate (*\label{line:cand}*)
  on partition<elect>(All, 1) (*\label{line:elect}*)
    win:  goto Leader
    lose: goto Replica (*\label{line:electe}*)
location Leader
  on recv(doCmd) do (*\label{line:request}*)
    cmd (*$\coloneqq$*) doCmd.payld
    if(cmd = 3) goto Return
    else goto RepCmd
...
\end{sudsl}
\end{tcolorbox}
%\caption{A partial process definition}
%\label{fig:syntax} 
%location RepCmd
%  on consensus<vcCmd>(All,1,cmd) do (*\label{line:cons1}*)
%    cmd (*$\coloneqq$*) vcCmd.decVar[1] (*\label{line:dec1}*)
%    ...
%location Replica
%  on consensus<vcCmd>(All,1,_) do (*\label{line:cons2}*)
%    cmd (*$\coloneqq$*) vcCmd.decVar[1] (*\label{line:dec2}*)
\end{wrapfigure}
\noindent system is composed of an arbitrary number of 
$n$ identical \aml system processes with process identifiers $1,\ldots,n$ and one environment process. The programmer specifies a system process definition $\process$ that consists 
of 
%(i) a set of typed, domain-bounded \emph{local variables},
(i) a set $\vars$ of \emph{local variables} with finite domains,
 (ii) a set $\events$ of \emph{events}
%declarations
used to communicate between processes, and (iii) a set of \emph{locations}
%, each containing a set of event handlers, 
that the processes can move between. 
Each event $\event$ in $\events$ incarnates an \emph{acting action} $\acting{\event}$ and a \emph{reacting action} $\reacting{\event}$ (e.g., for a \pairwise event, the acting (resp. reacting) action is the send (resp. receive) of that event). 
All processes start in a location denoted \terma{initial}.
Each location contains a set of \emph{action handlers} a process in that location can execute. Each handler has an associated action, a Boolean guard over the local variables, and a set of update statements. 
A partial process definition is depicted on the right. 

%\editt{add a sentence relating events to actions in a way.} \editt{add the sentence about locations and handlers here.In general, pull all the syntax up here?} 
%\edit{super shrink this.}

%\edit{move elsewhere or remove}
%In addition to the user-defined local variables,
%we assume the existence of a special ``location variable'' $\locVar$ which stores the current location of the process. 
%\edit{move to the end}
%We denote by $\vars$ the set of all local variables.
%Each local variable is either of type \terma{idSet}, indicating that the values it can hold are sets of process identifiers, or of type \terma{int[}\termb{l}\terma{,}\termb{u}\terma{]} indicating that the values it can hold are within the finite integer range $[\termb{l},\termb{u}]$. To simplify presentation, we assume the existence of a special ``location variable'' $\locVar$ which stores the current location of the process. 
%We denote by $\vars$ the set of all local variables.

The language supports five different types of events, namely, broadcast, \pairwise, partition, consensus, and internal.
The \emph{synchronous} broadcast (resp. \pairwise) \emph{communication} event type is denoted \terma{br} (resp. \terma{rz}) and indicates an event where one process synchronously communicates with all other processes (resp. another process).
%
%
%Each event declaration includes an event name, an event type, and an optional payload type. There are five different event types, namely, broadcast, \pairwise, partition, consensus, and internal. 
%The broadcast (resp. \pairwise) \emph{communication} event type is denoted \terma{br} (resp. \terma{rz}) and indicates an event where one process communicates with all other processes (resp. another process).\mn{skipped payload type description} 
%Both communication events allow for a payload type that is either \terma{unit} if no payload is needed or \terma{int[}\termb{l}\terma{,}\termb{u}\terma{]} indicating that the value being communicated in is within the finite integer range $[\termb{l},\termb{u}]$. 
%They can additionally include the keyword \terma{env} if they are a communication with the environment process.
%
%
%
%The \emph{agreement} event type partition %, denoted \terma{partition}, 
%indicates an event where a set of processes agree to partition themselves into winners and losers. The \emph{agreement} event type consensus
%%, denoted \terma{consensus}, 
%indicates an event where a set of processes reach consensus on a given set of decided values. Finally, the internal event
%%, denoted \terma{in}, 
%indicates an event where a process is performing its own internal computations. We denote by $\events$ the set of all communication and agreement events in the system. 
The \emph{agreement} event type partition, denoted \terma{partition}, 
indicates an event where a set of processes agree to partition themselves into winners and losers. For instance, in the figure, \terma{partition}\termb{<elect>} \termb{(All,1)} denotes a leader election round with identifier \termb{elect} where \termb{All} processes elect \termb{1} winning process that moves to the \termb{Leader} location, while all other losing processes move to the \termb{Replica} location.
The \emph{agreement} event type consensus, denoted \terma{consensus}, 
indicates an event where a set of processes, each proposing one value, reach consensus on a given set of decided values. For instance, \terma{consensus}\termb{<vcCmd>(All,1,cmd)} denotes a consensus round with identifier \termb{vcCmd} where \termb{All} processes want to agree on \termb{1} decided value from the set of proposed values in the local variable \termb{cmd}. 
%Upon termination of agreement, all processes retrieve the agreed-upon update to execute on the stored data using the \termb{vcCmd.}\terma{decVar}\termb{[1]} expression. 
Finally, the internal event
%, denoted \terma{in}, 
indicates an event where a process is performing its own internal computations. 
%We denote by $\events$ the set of all 
%communication and agreement 
%events in the system. 
%Communication and agreement e
%Each event $\event$ in $\events$ incarnates an \emph{acting action} $\acting{\event}$ and a \emph{reacting action} $\reacting{\event}$. 
For a communication event, the acting action is a send, while the reacting action 
is a receive. 
For a partition event, the acting action is a win, while the reacting action is a lose. Finally, for a consensus event, the acting action is proposing a winning value, while the reacting action is proposing a losing value. 
We denote by $\acting{\events}$ and $\reacting{\events}$ the set of all acting and reacting actions, respectively.

%Events in $\events$ incarnate a set of \emph{actions}. An action is 
%effectively 
%an event with a polarity. There are two polarities: acting and reacting. For some event $\event \in \events$, we denote the acting action as $\acting{\event}$ and the reacting action as $\reacting{\event}$. For a communication event, the acting action is a send, while the reacting action 
%is a receive. 
%For a partition event, the acting action is a win, while the reacting action is a lose. Finally, for a consensus event, the acting action is proposing a winning value, while the reacting action is proposing a losing value. We denote by $\acting{\events}$ and $\reacting{\events}$ the set of all acting and reacting actions, respectively.

%\edit{move elsewhere?}
The updates in an action handler may contain send, assignment, goto, and/or conditional statements. Assignment statements are of the form \termb{lhs $\coloneqq$ rhs} where \termb{lhs} is a local variable and \termb{rhs} is an expression of the appropriate type.
The goto statement \terma{goto} $\ell$ causes the process to switch to location $\ell$ (i.e., it can be thought of as the assignment statement $\locVar \coloneqq \ell$, where $\locVar$ is a special ``location variable'' that stores the current location of the process). 
The conditional statements are of the expected form: \terma{if(}\termb{cond}\terma{) then...else...}. We denote by $\handlers$ the set of all handlers in the process, and for each handler $\handler \in \handlers$ we denote its action, guard, and updates as $\handlerAction(\handler)$, $\guard(\handler)$, and $\update(\handler)$, respectively.\ourskip %, and/or \terma{goto} statements that are used to switch to a different location. % handlers

%In what follows we will define the local and global semantics of a \aml system.\ourskip

%\para{Local Semantics of a \aml Process.} 
\para{Local Semantics.} The local semantics $\semantics{\process}$ of a process $\process$ is expressed as a state-transition system $(\LQ,\loq_0,\events,\T)$, where $\LQ$ is the set of local states, $\loq_0$ is the
initial state, $\events$ is the set of 
events, 
and $\T \subseteq \LQ \times \{ \acting{\events} \cup \reacting{\events} \} \times \LQ$ is the set of transitions of $\semantics{\process}$.
%and $\T \subseteq \Ts$ is the set of local labeled transitions of $P$. Here, $\Ts = \LQ \times \eventPolarity \times \eventTypes \times \events \times \LQ$ where:
%\begin{compactenum}[-]
%\item $\eventTypes = \{ \mathtt{rz},\mathtt{br},\mathtt{partition},\mathtt{consensus}, \epsilon\}$ is the set of action types ....
%\item $\eventPolarity= \{Acting , Reacting\}$ is the set of action polarities ....%$\mathbf{\mathscr{C}}$
%\end{compactenum}
A state $\loq \in \LQ$
is a valuation of the variables in $\vars$.
% as well as the location variable $\locVar$ indicating the current location of the process. 
We let $\valOf{\ls}{\var}$ denote the value of the variable $\var$ in state $\ls$.
%Similarly, we use $\valOf{\ls}{expr}$ to denote the value of 
%expression $expr$ evaluated in state $\ls$.

%For a state $\loq = (\q,\sigma)$, we let $\loq.loc$ denote the location $\q$ in $\loq$, and 
%$ \valOf{\loq}{var}$ denote the value $\sigmaOf{var}$ of variable $var$ in $\loq$. 
%The initial state $\ls_0 = (\q_0, \sigma_0)$, where $\q_0$ denotes the initial location and $\sigma_0$ denotes the initial variable valuation. 
The set of action handlers associated with all acting and reacting actions of all events induces the transitions in $\T$. In particular,
%to the execution of one of the core event handlers of \aml; a transition 
%is labeled either with a send/receive of a communication 
%action in $acts$, an empty label $\epsilon$ denoting an internal transition, or one of the two agreement primitives.
a transition $t = \trans{\ls}{\action}{\ls'}$ based on action handler $\handler$ over action $\action$ is in $\T$ iff the guard $\guard(\handler)$ evaluates to $true$ in $\ls$ and $\ls'$ is obtained by applying the updates $\update(\handler)$ to $\ls$.\ourskip

%\para{Global Semantics of a \aml System.}
\para{Global Semantics.} The global semantics $\globalSemantics{\process}{n}$ 
%\mn{drawback of using this is that we need to define $\system$ everywhere we use it?}
%\nour{use [[$P_n$]] and for the syntax use p1 || ... pn} 
of a \aml system $\process_1 || \ldots$ $|| \process_n || \process_e$
%$\system$ 
consisting of $n$ identical 
%\neww{ , synchronously-communicating} 
processes $\process_1, \ldots,\process_n$ and an environment process $\process_e$ (with local state space $\LS_e$)
%, communicating synchronously, 
is expressed as a transition system $(\GQ,\gq_0,\events,\GT)$, where $\GQ = \LQ^n \times \LQ_e$ is the set of global states, $\gq_0$ % = (\ls_0, \ldots, \ls_0, \ls_{0,e})$ 
is the
initial global state, $\events$ is the set of 
events,
and $\GT \subseteq \GQ \times \events \times \GQ$ is the set of global transitions of $\globalSemantics{\process}{n}$.

The set of events $\events$ induce the transitions in $\GT$. As is the case for events, there are five types of global transitions: broadcast, \pairwise, partition, consensus, and internal.
%(i) a broadcast transition where one process communicates with all other processes, (ii) a \pairwise transition where one process communicates with another, (iii) a partition transition where a set of processes partition themselves into winners and losers, (iv) a consensus transition where a set of processes decide on a subset of values from a set of proposed values, and (v) an internal transition where one process performs an internal computation. 
In particular, a transition $r = \trans{\gq}{\event}{\gq'}$ for some broadcast event $\event$ is in $\GT$ iff the send local transition $\trans{\gq[i]}{\acting{\event}}{\gq[i]'}$ is in $\T$ for some process $\process_i$, and the receive local transition  $\trans{\gq[j]}{\reacting{\event}}{\gq[j]'}$ is in $\T$ for every other process $\process_j$ with $j \neq i$. The remaining global transitions can be formalized similarly. 
%We refer the interested reader to \cite{Jaber.QuickSilver.OOSLA.2021} for a detailed treatment of the local and global semantics of \aml processes and systems.

A trace of a \aml system is a sequence $\gq_0, \gq_1, \ldots$ of global states such that for every $i \geq 0$, the global transition $\trans{\gq_i}{\event}{\gq_{i+1}}$ for some event $\event$ is in $\GT$. A global state $\gq$ is \emph{reachable} if there is a trace that ends in it. \ourskip

\para{Permissible Safety Specifications.} \kinarach targets parameterized verification for a class of properties called \emph{permissible} safety specifications that disallow global states where $m$ or more processes, for some fixed number $m$, are in some subset of the local states. We denote by $\safetySpec$ the permissible safety specifications provided by the designer for a system with $n$ processes.
A \aml system is safe if there are no reachable error states in its global semantics. We denote that as $\globalSemantics{\process}{n} \models \safetySpec$.\ourskip

\para{The \EFfdecidable Fragment.}
%In order to solve the parameterized verification problem for \aml systems, 
\kinarach identifies a fragment of \aml for which the parameterized verification problem of a large class of safety properties is \emph{\effdecidable}. 
	In particular, a pair $\langle \process,\phi\rangle$ of a \aml process $\process$ and a safety specification $\phi$ is in the \effdecidable fragment of \aml if it satisfies \emph{\chwellbehavedness} and \emph{\amenability} conditions. For such a pair, a \emph{cutoff} number $c$ of processes can be computed and the parameterized verification problem can be reduced to the verification of the cutoff-sized system (i.e., $\forall n: \globalSemantics{\process}{n} \models \safetySpec \Leftrightarrow \globalSemantics{\process}{c} \models \safetySpecc$).
	
	During verification, \kinarach computes a set of \emph{phases} that an execution of the system goes through. 
	On a high level, the \chwellbehavedness conditions ensure that the system moves between phases through ``globally-synchronizing'' events (i.e., broadcast, partition, or consensus), and that all processes in the same phase can participate in further globally-synchronizing events. 
	This ensures that the system's ability to move between phases is \emph{independent} of the number of processes in the system. The \amenability conditions ensure that an error state, where $m$ processes are in a subset of the local states violating some 
	%permissible 
	safety specification, is reachable in a system of any size iff it is reachable in a system with exactly $m$ processes. 
	%When a pair $\langle \p,\phi\rangle$ meets 
	%these conditions, the parameterized verification of a system composed of an arbitrary number $n$ of processes can be reduced to the verification of a system with a fixed, finite number $c$ of processes. 
If any of these conditions fails, the designer must modify the process definition manually and attempt the verification again. We denote by $\semantics{\process} \models \phaseCompSpec$ (resp. $\semantics{\process} \models \cutoffsSpec$) that the \aml process $\process$ with local semantics $\semantics{\process}$ satisfies \chwellbehavedness (resp. \amenability) conditions. 
%\editt{if needed add something like the nature of these properties (FOL) formulas etc, or we can wait on that for the extraction?}
	%We will refer to such conditions as \emph{local properties}.
	%\para{C.} .... .We will refer to such conditions as \emph{global properties}.
%	\edit{Local Properties needed for decidability} talk about phase comp and amenability here.
%	\nour{text from Notion:}
%	
%	**Correctness Properties ($prop$)**
%	
%	The correctness property $prop$ should simply be a FOL formula/predicate over the local semantics $LS$. We require that $LS \models prop$. (or $prop(LS) = true$).
%	
%	QuickSilver is a model checker: given a specific $LS$ and $prop$ it returns $true$ if $LS \models prop$ and $false$ otherwise. 
%	
%	**Local Semantics ($LS$)**
%	\edit{refer the read to quicksilver for bla bla..}
%	
\subsection{\aml Process Sketch}
\label{sec:sketches}
Let us extend \aml's syntax to allow process \emph{sketches} that can be completed by a synthesizer. 
In particular, we allow the process definition to include a set 
%$\uFuns$ 
of \emph{uninterpreted functions} that can replace various expressions in \aml such as the Boolean expression \termb{cond} in the \terma{if(}\termb{cond}\terma{) then} $\ldots$ \terma{else} $\ldots$, the target locations of \terma{goto} statements, and the \termb{rhs} of assignments.
\footnote{Such uninterpreted functions are sufficient to be a building block for more complex expressions and statements (See, for instance, the \textsc{Sketch} Language \cite{cegis}). 
		%For instance, the boolean uninterpreted function $\uf$ in \terma{if(}$\uf$\terma{)} $if_{body}$ \terma{else} \termb{Logic2} can be used to model selection of the logic to incorporate in place of this conditional.
	}
%; similar to how Sketch \cite{cegis} builds on one integer hole. \edit{add justification}
%Currently, \tool only supports sketches with unspecified expressions. While this has the advantage of guaranteeing termination, it requires the user to use unspecified expressions as building blocks to synthesize more complex structures (e.g., using \terma{if(}$\ufDSL{1}$\terma{) goto} \termb{Logic1} \terma{else goto} \termb{Logic2} to model choosing which logic to incorporate in place of this conditional).
As is standard, each uninterpreted function $\genericUFun$ is equipped with a signature determining its list of named, typed parameters and its return type. 
%We define function arguments and applications in the standard way. 
A valid list of arguments $\argument$ for some function $\genericUFun$ is a list of values with types that match the function's parameter list. Applying a function $\genericUFun$ to a valid list of arguments $\argument$ is denoted by  $\genericUFun(\argument)$.
 Additionally, we define a \emph{function interpretation} $\interpretation(\genericUFun)$ of an uninterpreted function $\genericUFun$ as a mapping from every valid list of arguments of $\genericUFun$ to a valid return value.
  
A \aml process definition $\process$ that contains one or more uninterpreted functions is called a \emph{sketch}, and is denoted $\psketch$. 
We denote by $\uFuns$ the set of all uninterpreted functions in a sketch $\psketch$. An interpretation $\interpretation$ \mn{While subscripting $I$ with ``sk'' is nicer here, it'll cause a lot of headache for the rest of the formalism as $I$ is a subscript almost everywhere. We will skip that and explicitly say what $I$ is over. Alternatively, we can put back the set $\interpretations$ which was subscripted.} of the set $\uFuns$ of uninterpreted functions is then a mapping from every uninterpreted function $\uFun \in \uFuns$ to some function interpretation $\interpretation(\uFun)$. 
%We denote by $\interpretations$ the set of all possible interpretations 
%for all uninterpreted functions induced by $\psketch$. 

%\ourskip
%\para{Interpreted Process Sketch.} 
%For some interpretation $\interpretation$ and some process sketch $\psketch$,
For some process sketch $\psketch$ and some interpretation $\interpretation$ of the set $\uFuns$ of uninterpreted functions in $\psketch$, we denote by
$\pcomplete$ the \emph{interpreted process sketch} obtained by replacing every uninterpreted function $\uFun \in \uFuns$ in the sketch $\psketch$ with its function interpretation $\interpretation(\uFun)$ according to the interpretation $\interpretation$. 

\subsection{Problem Definition}
\label{sec:problemDef} 

%In this section, 
We now define the parameterized synthesis problem for \aml systems. 
\begin{definition} [\aml Parameterized Synthesis Problem (\mpsp)]
Given a process sketch $\psketch$ with a set of uninterpreted functions $\uFuns$, an environment process $\process_e$, and \permissible safety specification $\safetySpec$, find an interpretation $\interpretation
% \in \interpretations
 $ of uninterpreted functions in $\uFuns$ such that the system $\pcompletee{1} || \ldots || \pcompletee{n} || \process_e$
%$\system$, consisting of processes $\pcompletee{1}, \ldots, \pcompletee{n}$ and $\process_e$,}
%interpreted process sketch $\pcomplete$ 
is safe for any number of processes, \ie, 
%\coloneq \interpretation(\psketch)
%$ with a corresponding global semantics $\iglobalSemantics(n) \coloneqq {\pcomplete}_1 || \ldots || {\pcomplete}_n || \p_e$ such that 
$\forall n: \iglobalSemantics{\pcomplete}{n} \models \safetySpec$. %\editt{see the commented, more-nuanced version}
\end{definition} 
\section{Constraint-Based Synthesis for \aml Systems}
\label{sec:algorithm}

\begin{figure}[t]
%\squeezecaption
%\squeezecaption
%\squeezecaption
%\squeezecaption
%\squeezecaption
%\squeezecaption
\centering
\includegraphics[width=0.8\linewidth]{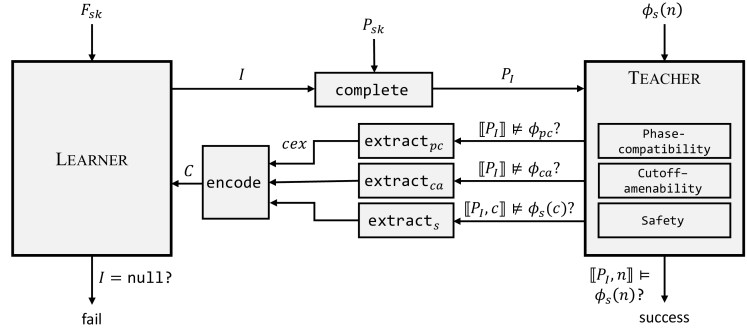}
\caption{Overview of \tool's architecture.}
% In the figure, $pc$ and $ca$ are short for \chwellbehavedness and \amenability, respectively.
%}
\label{fig:flow}
%\squeezecaption
%\squeezecaption
%\squeezecaption
%\squeezecaption
%\squeezecaption
%\squeezecaption
%\squeezecaption
%\squeezecaption
%\squeezecaption
%\squeezecaption
\end{figure}
%\nour{highlight parameterized safty vs liveness}
\para{Architecture.} To solve \mpsp, we propose a multi-stage, counterexample-based architecture, shown in \figref{flow}, with the following components: 
%\mn{The dashes are the default format and are used in previous TACAS papers.}
\begin{compactitem}
\item \textsc{Learner}: 
  %, denoted $\learnerSymbol$, based on the Z3 SMT Solver \cite{z3} 
a constraint-solver that accepts a set $\constraints$ of \emph{constraints} over the uninterpreted functions $\uFuns$ and generates interpretations $\interpretation$ satisfying these constraints (i.e., $\interpretation \models \constraints$). Specifically, a constraint $\constraint \in \constraints$ is a well-typed Boolean formula over uninterpreted function applications. 
%and values in the corresponding functions' return types. 
  %Effectively, every constraint the \learner receives serves as a ``hint'' on what the correct interpretation should be. Note that if the \learner runs out of interpretations to provide, then the synthesis procedure has failed.} %Note that the oracle 

  \item \textsc{Teacher}: 
%\cite{Jaber.QuickSilver.OOSLA.2021} 
%is a \teacher for 
%\kinarach, 
a component capable of checking \chwellbehavedness, \amenability, safety, and \emph{liveness}\footnote{
While \mpsp targets permissible safety specifications, in order to improve the \emph{quality} of the interpreted process sketch $\pcomplete$, we extend \aml with \emph{liveness} specifications to 
%, denoted $\livenessSpec$ where $k$ is a {\em fixed} number of processes. Such specifications 
help rule out trivial completions that are safe. We emphasize that such specifications are only used as a tool to improve the quality of synthesis, and are only guaranteed for the cutoff-sized system, as opposed to safety properties that are guaranteed for any system size. 
%We elaborate on liveness properties in \secref{implementation}.
} of \aml systems. We refer to these four conditions as \emph{properties}.
%\nour{add a footnote here and move all liveness stuff to implementation. Hide from algo and arch?} \nour{add a remark about ``properties'' in general}
%capable of checking \chwellbehavedness and \amenability conditions. %As discussed in \secref{sketches}, we extend \kinarach to accept process sketches and keep track of disabled local transitions in the local semantics.  

\item \code{complete}: a component that builds an interpreted process sketch $\pcomplete$ from a process sketch $\psketch$ and an 
  %by replacing every uninterpreted function $\uFun \in \uFuns$ in $\psketch$ with its function interpretation $\interpretation(\uFun)$ according to the 
  interpretation $\interpretation$ provided by the \learner.

\item 
\code{extract$_{prop}$}: 
%\nour{make this parmaterized over the properties}
a \emph{property-specific} component to extract a counterexample $\cex$, capturing the root cause of a violation, if the \textsc{Teacher} determines that a property $prop$ from the above-mentioned properties is violated.

\item \code{encode}: a novel \emph{property-agnostic} component that encodes counterexamples generated by \code{extract} components into additional constraints 
 for the \learner.
% An encoding of some counterexample $\cex$ ensures that \emph{any} interpretation provided by $\learnerSymbol$ will not exhibit a similar violation to that captured by $\cex$. We expand on the encoding procedure in \secref{encoding}.
\end{compactitem}
\ourskip

\begin{wrapfigure}[22]{r}{0.51\textwidth}
\vspace{-2.5em}
    \begin{minipage}{0.515\textwidth}
    \footnotesize
 \SetInd{0.28em}{0.43em}
 \begin{algorithm}[H]
   \SetKwFunction{synthisProc}{Synth}  
   \SetKwProg{myproc}{procedure}{}{}
   \myproc{\synthisProc{$\psketch, \safetySpec, \livenessSpec$}}{
   \SetKwInOut{Input}{Input}
   \SetKwInOut{Output}{Output}
 %  \Input{A \aml program sketch, $\psketch$}  
 %  \Input{\Permissible safety specification, $\safetySpec$}
 %  \Input{Liveness specification, $\livenessSpec$}
 %  \Output{A \chwellbehaved, \amenable, safe, and live  process $\pcomplete$, or \code{null}}
 % \BlankLine
  
   $\constraints = \varnothing$ \label{line:emptyconstraints}%\Comment{Empty set of constraints} 
      	     
   \While{true}{
     $\interpretation$ = \Call{interpret}{$\uFuns,\constraints$} \label{line:interpret} %\Comment{Obtain an interpretation from a \learner}
 	
     \If{$\interpretation \neq$ \code{null}} {
 
       $\pcomplete = $ \Call{complete}{$\psketch, \interpretation$} \label{line:completed}
       
       $\isemantics{\pcomplete} = $ \Call{buildLS}{$\pcomplete$}
       
       $\cexPhComp$ = \Call{findPhCoCE}{$\isemantics{\pcomplete}$}
       \label{line:compatibility} %\Comment{A \counterexample to \chwellbehavedness}

       \If(\label{line:notcompatible} 
       %\Comment{$\pcomplete$ is not \chwellbehaved}
       ){$\cexPhComp  \neq$ \code{null}}{

        	$\constraints = \constraints \ \cup \neg$ \Call{encode}{$\cexPhComp$} \label{line:encodeph}%\Comment{Add constraint to eliminate $\cexPhComp$}
 
         \textbf{Continue} \label{line:startoverp}%\Comment{Start over}
       }       
% 	  \BlankLine
% 
%       $\cexCutoff$ = \Call{findCutoffAmenabilityCE}{$\isemantics, \safetySpec$}
%       \label{line:amenability} %\Comment{A \counterexample to \amenability}
%       
%       \If(\label{line:notamenable} 
%       %\Comment{$\pcomplete$ is not \amenable}
%       ){$\cexCutoff  \neq$ \code{null}}{
% 		
% 		
%        	$\constraints = \constraints \ \cup$ \Call{encode}{$\cexCutoff$} \label{line:encodeca}%\Comment{Add constraint to eliminate $\cexCutoff$}
% 
%         \textbf{Continue} %\Comment{Start over}
%       }       
 	  ... $\rhd$ check \amenability
 	  
 	 % \BlankLine
       
       $c$ = \Call{compCutoff}{$\pcomplete, \safetySpec$} \label{line:cutoff}
       
       $\iglobalSemantics{\pcomplete}{c} = $ \Call{buildGS}{$\pcomplete,c$}

       $\cexSafety$ = \Call{findSaCE}{$\iglobalSemantics{\pcomplete}{c},\safetySpecc$}
       \label{line:safety} %\Comment{A \counterexample to safety}
       
       \If(\label{line:notsafe} 
       %\Comment{$\pcomplete$ is not safe}
       ){$\cexSafety  \neq$ \code{null}}{

        	$\constraints = \constraints \ \cup \neg$ \Call{encode}{$\cexSafety$} \label{line:encodes}%\Comment{Add constraint to eliminate $\cexSafety$}
 
         \textbf{Continue} %\Comment{Start over}
       }       
% 	  \BlankLine
% 
%       $\cexLiveness$ = \Call{findLivenessCE}{$\iglobalSemantics, \livenessSpec$}
%       \label{line:liveness} %\Comment{A \counterexample to liveness}
%       
%       \If(\label{line:notlive} 
%       %\Comment{$\pcomplete$ is not live}
%       ){$\cexLiveness  \neq$ \code{null}}{
% 		
% 		
%        	$\constraints = \constraints \ \cup$ \Call{encode}{$\cexLiveness$} \label{line:encodel}%\Comment{Add constraint to eliminate $\cexLiveness$}
% 
%         \textbf{Continue} %\Comment{Start over}
%       }       
% 	  \BlankLine
%             
%... $\rhd$ check liveness

       \Return $\pcomplete$ %\Comment{Found a correct completion!}
       		
 	}\Else{
 	  \Return  \code{null} %\Comment{Synthesis failed, break and return \code{null}}
 	}
 	
   } % end of while true
 } % end of procedue
 \caption{
 %Procedure for 
 Solving \mpsp.}
 \label{algo:synthAlgo}
 \end{algorithm}
    \end{minipage}
  \end{wrapfigure}

\para{Synthesis Procedure.} 
\tool instantiates this architecture as shown in \algoref{synthAlgo}. 
%The algorithm first checks if an interpreted process sketch satisfies \chwellbehavedness and \amenability, and then proceeds to compute a cutoff value, and check if the cutoff-sized system is safe and live. More precisely, t
The algorithm starts with an empty set of constraints, $\constraints$ (\lineref{emptyconstraints}) over the set $\uFuns$ of uninterpreted functions in the process sketch $\psketch$.
In each iteration, it checks if there exists an interpretation $\interpretation$ of the uninterpreted functions that satisfies all the constraints collected so far (\lineref{interpret}). 
If such an interpretation is found, it is used to obtain an interpreted process sketch $\pcomplete$ (\lineref{completed}). 
Then, the algorithm checks if the system described by $\pcomplete$ is \chwellbehaved and \amenable. If so, a cutoff $c$ is computed (\lineref{cutoff}) and the $c$-sized system is checked to be
safe. 
%and live. 
The \amenability  stage
%and liveness stages are 
is similar to \chwellbehavedness 
%and safety stages, respectively, and are
and is hence omitted from the algorithm.
 At any stage, if the process fails to satisfy any of these properties (e.g., a counterexample $\cexPhComp$ to \chwellbehavedness is found on \lineref{compatibility}), the root-cause of the failure is extracted and encoded into a constraint for the \learner to rule out the failure (e.g., \lineref{encodeph}). 

%\tool instantiates this architecture according to the algorithm  in \algoref{synthAlgo}.
%The procedure first checks if an interpreted process sketch satisfies \chwellbehavedness and \amenability, and then proceeds to 
%the latter two stages where it 
%compute a cutoff value, and check if the cutoff-sized system is safe and live for $k$.
%More precisely, t
%The procedure starts 
%We with an empty set of constraints, $\constraints$ 
%(\lineref{emptyconstraints}) 
%over the set $\uFuns$ of uninterpreted functions in the process sketch $\psketch$.

%In each iteration, the procedure attempts to obtain an interpretation $\interpretation$ of the uninterpreted functions that satisfies all the constraints collected so far. %(\lineref{interpret}). 
%If such an interpretation exists, the interpretation is used to obtain an interpreted process sketch $\pcomplete$. %(\lineref{completed}). 
%Then, the procedure checks if the system described by $\pcomplete$ is \chwellbehaved, \amenable, safe, and live. At any stage, if the process fails to satisfy any of these properties, the root-cause of the failure is extracted and encoded into a constraint for the \learner to rule out the failure. 

%\mn{addressing why are stages sequential}
Note that these stages are checked sequentially due to the inherent dependency between them: (i) the system can only be cutoff amenable if it is phase compatible, and (ii) one can only check safety after a cutoff has been computed.

\begin{lemma}
\label{lem:sound}
%Assuming that the \teacher is sound and the \learner is complete, 
%%for finite sets of interpretations, 
%the above procedure for solving \mpsp is sound and complete. 
Assuming that the \teacher is sound and the \learner is complete for finite sets of interpretations, \algoref{synthAlgo} for solving \mpsp is sound and complete. 
\end{lemma}
\noindent{\em Proof.}
Soundness follows directly from the soundness of the \teacher. 
%The soundness of the procedure follows directly from the soundness of \kinarach, shown in \cite{Jaber.QuickSilver.OOSLA.2021}. In particular, The first two stages ensure that the \aml Parameterized Verification Problem is decidable for the candidate process $\pcomplete$ and that $c$ is valid cutoff (i.e., $\iglobalSemantics(c) \models \safetySpecc \Leftrightarrow \forall n:\iglobalSemantics(n) \models \safetySpec$). Stage three ensures that the cutoff-seized system is indeed correct (i.e., $\iglobalSemantics(c) \models \safetySpecc$). The last stage to check liveness merely filters out trivial solutions and hence, has no effect on soundness.
Completeness follows from that the encoding and extraction procedures ensure progress by eliminating at least the current interpretation at each iteration, and the finiteness of the set 
of interpretations.  
%Then, the loop will either terminate with a correct candidate process $\pcomplete$, or exhaust all possible interpretations and terminate with \code{null}. 
Finiteness follows from (i) the finite number of uninterpreted functions in a sketch $\psketch$, (ii) the finiteness of the domain of each local variable, 
and (iii) the finiteness of the number of local variables in $\psketch$.\ourskip

In the remainder of this section, we describe the property-agnostic \code{encode} component in \algoref{synthAlgo}. 
In the following section, we describe our implementation of %\algoref{synthAlgo} 
our synthesis procedure specialized to a \kinarach-based \teacher and 
property-specific extraction procedures.

\subsection*{Property-Agnostic \Counterexample Encoding Procedure}
\label{sec:encoding}
%\nour{edit the section to make this a procedure}
 %\mn{Semantics extensions moved to solution} 
 We first describe the necessary augmentation of local 
 %and global 
 semantics with \emph{disabled transitions} needed for \tool's counterexample extraction and encoding. While such transitions are not relevant when reasoning about a ``concrete'' process definition (i.e., one with no uninterpreted functions), they are quite important when extracting an \emph{explanation} for why some conditions (e.g., \chwellbehavedness) fail to hold on $\isemantics{\process}$.\ourskip

\para{Augmented Local Semantics of the \aml Process $\mathbf{\pcomplete}$.} We extend the definition of the local semantics of a \aml interpreted process sketch $\pcomplete$ to be $\isemantics{\pcomplete} = (\istates,\loq_0,\events,\itransitions, \idistransitions)$ where $\istates$, $\loq_0$, $\events$, and $\itransitions$ are defined as before and $\idistransitions$ is the set of \emph{disabled transitions} under the current interpretation $\interpretation$. In particular, a disabled transition $t = \trans{\ls}{\action}{\bot}$ based on action handler $\handler$ over action $\action$ is in $\idistransitions$ iff the guard $\guard(\handler)$ evaluates to $false$ in $\ls$. The symbol $\bot$ here indicates that no local state is reachable, since the guard is disabled.
%\ourskip
%\paraa{Concrete Local States.} 

Additionally, we say a transition $t = \trans{\ls}{\action}{\ls'}$ based on action handler $\handler$ over action $\action$ is a \emph{sketch transition} if $\handler$ contains no uninterpreted functions in its guard or updates. A local state $\ls \in \istates$ is \emph{concrete} if (i) $\ls$ is the initial state $\ls_0$, or
(ii) there exists a sketch transition $\trans{\ls'}{}{\ls}$ where $\ls'$ is concrete. In other words, a local state $\ls$ is concrete if there exists a path from the initial state $\ls_0$ to $\ls$ that is composed purely of sketch transitions and hence is always reachable regardless of the interpretation we obtain from the \learner.\ourskip

%\neww{For instance, a disabled global transition $r = \trans{\gq}{\event}{\bot}$ based on broadcast event $\event$ is in $\iDisGT$ iff (i) there exists a process $\p_{\interpretation,i}$ with 
%$\trans{\gq[i]}{\acting{\event}}{\gq[i]'}$ is in $\itransitions$ but for some other process $\p_{\interpretation,j}$ we have $\trans{\gq[i]}{\reacting{\event}}{\bot}$ is in $\idistransitions$, or (ii) for all processes $\p_{\interpretation,i}$ in $\gq$, we have $\trans{\gq[i]}{\acting{\event}}{\bot}$ is in $\idistransitions$. Other disabled transitions are defined similarly.}
 %set of processes $\p_{\interpretation,i},\ldots,\p_{\interpretation,m}$ in $\gq$ that are unable to take part of the event.}
% 
% 
% $\trans{\gq[i]}{\acting{\event}}{\gq[i]'}$ is in $\T$ for some process $\p_i$, and the receive local transition  $\trans{\gq[j]}{\reacting{\event}}{\gq[j]'}$ is in $\T$ for every other
% 
% such that 
% 
% the guard $\guard(\handler)$ evaluates to $false$ in $\ls$. 

%Our synthesis procedure (ref. \algoref{synthAlgo}) for solving \mpsp hinges on extracting \counterexamples for when ``properties'' fail at any stage of the synthesis loop, and encoding such \counterexamples as constraints to the \learner so that all interpretations exhibiting the same violation are ruled out. For the reminder of this paper, we will refer to \chwellbehavedness and \amenability conditions as \emph{local properties} and safety and liveness specifications as \emph{global properties}.

We now formalize \counterexamples for 
\chwellbehavedness and cutoff amenability properties 
then present an encoding procedure for such 
\counterexamples. 
%On a high level, the encoding of a \counterexample $\cex$ corresponding to a local or a global property violation yields a constraint $\constraint$ that can be added to the \learner $\learnerSymbol$ to rule out that violation. 
The encoding is {\em exact} in the sense that a generated constraint $\constraint$ corresponding to some \counterexample $\cex$ rules out exactly all interpretations $\interpretation
 %\in \interpretations
 $ 
where an interpreted process sketch 
%a candidate process %local or global semantics of 
$\pcomplete$ exhibits $\cex$ (as opposed to an over-approximation where $\constraint$ would rule out interpreted process sketches that do not exhibit $\cex$, or an under-approximation where $\constraint$ would allow interpreted process sketches that do exhibit $\cex$). Additionally, the encoding is {\em property-agnostic} in the sense that it can handle \counterexamples for any property failure. \ourskip
%\subsection{Local Properties}
%\Chwellbehavedness and \amenability conditions ($\phaseCompSpec$ and $\cutoffsSpec$, respectively) are introduced in \cite{Jaber.QuickSilver.OOSLA.2021} as sufficient conditions over the local semantics of some \aml process $\p$ and permissible safety specifications $\safetySpec$ that render the parameterized verification problem efficiently decidable.
%%In general, such 
%These conditions are specified as first-order logic formulas over the local semantics of a \aml process. The exact set of conditions are detailed in \appref{extractionApp}. %\nour{or \cite{quicksilver}?}.

\para{\Counterexamples}. %of Local Properties}
Recall that a candidate process $\pcomplete$ based on some process sketch $\psketch$ and interpretation $\interpretation$ has the local semantics $\isemantics{\pcomplete} = (\istates,\loq_0,\events,\itransitions, \idistransitions)$. % where $\istates$ is the set of local states, $\loq_0$ is the initial state, $\events$ is the set of events and $\itransitions$ (resp. $\idistransitions$) is the set of enabled (resp. disabled) transitions under the current interpretation $\interpretation$. 
%Further, recall in the first two stages of the synthesis loop, we check if $\pcomplete$ satisfies \chwellbehavedness and \amenability conditions (i.e., $\isemantics \models \phaseCompSpec$ and $\isemantics \models \cutoffsSpec$).
A \counterexample $\cex$ to \chwellbehavedness (resp. \amenability) is a ``subset'' of the local semantics $\isemantics{\pcomplete}$ such that $\cex \not \models \phaseCompSpec$ (resp. $\cex \not \models \cutoffsSpec$). We say that $\cex$ is a subset of $\isemantics{\pcomplete}$, denoted $\cex \subseteq \isemantics{\pcomplete}$, when it has a subset of its enabled and disabled transitions, i.e., $\cex = (\istates,\loq_0,\events,\itransitions' \subseteq \itransitions, \idistransitionsPrime \subseteq \idistransitions)$.\ourskip

\para{Encoding \Counterexamples}. Let $\allConstraints$ be the set of all well-typed constraints that the \learner accepts. The encoding of \counterexample $\cex = (\istates, \loq_0, \events, \itransitions, \idistransitions)$ w.r.t. interpretation $\interpretation$ is a formula $\existsTr{\cex} \in \allConstraints$ defined as:
$$
\existsTr{\cex} = \Big(\bigwedge_{t_{en} \in \itransitions} \existsTr{t_{en}} \Big) \wedge \Big(\bigwedge_{t_{dis} \in \idistransitions} \existsTr{t_{dis}}\Big),
$$ 
where $\existsTr{t_{en}}$ (resp. $\existsTr{t_{dis}}$) is an encoding of an enabled (resp. disabled) local transition.
Note that $\existsTr{\cex}$
%An encoding of some enabled transition $t$ in $\cex$
%is a formula 
%$\fconstraint \in \allConstraints$ that
%over the atoms of the interpretation I that 
is satisfied under interpretation $\interpretation$ (i.e., $\interpretation \models \existsTr{\cex}$) and implies that $\cex \subseteq \isemantics{\process}$. %Informally, such a formula acts like an interpolant between the logical encoding of the interpretation $\interpretation$ and the logical encoding of the transition $t$.
An encoding of some enabled transition $t_{en} =  \trans{\ls}{\action}{\ls'}$
 based on action handler $\handler$ over action $\action$ is defined as: 
 $$
 %\fconstraint = 
 \existsTr{\trans{\ls}{\action}{\ls'}} = 
 \reachable{\ls} \wedge 
 \hasAction{\ls}{\action} \wedge 
  \goesTo{\ls}{\action}{\ls'},$$
  where:
%Given an interpretation $\interpretation$ and local semantics $\isemantics$ of completed process $\pcomplete$, a \emph{manifestation fingerprint} for some transition $t$ in $\isemantics$
%is a formula $\fconstraint \in \allConstraints$ that
%%over the atoms of the interpretation I that 
%(i) is satisfied under interpretation $\interpretation$ (i.e., $\interpretation \models \fconstraint$) and (ii) implies that $t \in \itransitions$. Informally, such fingerprint $\fconstraint$ acts like an interpolant between the logical encoding of the interpretation $\interpretation$ and the logical encoding of the transition $t$. In what follows, for some transition $t =  \trans{\ls}{\act}{\ls'}$, we show how to build the fingerprint $\fconstraint$, denoted $\existsTr{\trans{\ls}{\act}{\ls'}}$ as a conjunction of the following predicates:\ourskip
%elude to interpolants.
%A --> B --> C
%C (universe of the local semantics):
%TODO
%``partial eval''
%``good notation''
%A local state $\ls \in \istates$ is concrete if there exists a path from the initial state $\ls_0$ to $\ls$ in the \emph{program sketch} $\psketch$ \nour{can also define $\LS_{sk}$ and $\T_{sk}$ if need be.}. In other words, regardless of the interpretation we obtain from the oracle, $\ls$ is always reachable.\ourskip
\begin{compactenum}
\item the predicate $\reachable{\ls}$ indicating that the source state $\ls$ is reachable from the initial state $\ls_0$ under interpretation $\interpretation$. If $\ls$ is concrete, $\reachable{\ls}$ is simply $true$ (i.e., s is always reachable regardless of $\interpretation$). Otherwise, $\reachable{\ls}$ is defined as follows. Let $\mathcal{P}$ be the set of all paths from the initial state $\ls_0$ to state $\ls$. Then,
$\reachable{\ls} \coloneqq \bigvee_{p\in \mathcal{P}} \existsTr{p}$,
where $\existsTr{p}$ for some path $p$ consisting of local transitions $t_1,\ldots,t_i$ is defined as $\existsTr{t_1} \wedge \ldots \wedge \existsTr{t_i}$.

\item 
the predicate $\hasAction{\ls}{\action}$ indicating that the process can perform action $\action$ from state $\ls$. The predicate $\hasAction{\ls}{\action}$ is defined as follows:
%. 
%Let $\guard(h')$ be the guard associated with handler $h'$. Let $h$ be the enabled handler over action $\action$ in state $\ls$ (i.e. $\valOf{\ls}{\guard(h)} = true$ under interpretation $\interpretation$).\footnote{There will only be one enabled handler over a given action in a particular state due to determinism assumptions.\nour{add: a?? a?? on the same state and similarly a!! and a!!, check picture}} Then, 
$\hasAction{\ls}{\action} \coloneqq (\guard(\handler)[\valOf{\ls}{\vars}/\vars] = true)$,
%
%$$\hasAction{\ls}{\act} \coloneq (\guard(\handler)[\valOf{\ls}{\vars}/\vars] = \valOf{\ls}{\guard(\handler)}),$$
%\editt{double check this\footnote{\nour{revise this to be: collect ufs, make it uf.partialEval(src) = uf.eval(src)}}}
where $\guard(\handler)[\valOf{\ls}{\vars}/\vars]$ is the guard $\guard(\handler)$ with each local variable $v \in \vars$ replaced by its value $\valOf{\ls}{v}$ in state $\ls$.

%\item 
%%\nour{account for different handlers}
%the predicate $\hasAction{\ls}{\act}$ indicating that an action $\act$ is available in state $\ls$. The predicate $\hasAction{\ls}{\act}$ is defined as follows. Let $\handlers(\act)$ be the set of handlers over action $\act$, and $\guard(h)$ be the guard associated with handler $h$. Then, 
%$$\hasAction{\ls}{\act} = \bigvee_{h \in \handlers(\act)} \guard(h)[\valOf{\ls}{\V}/\V],$$
%where $\guard(h)[\valOf{\ls}{\V}/\V]$ is the guard $\guard(h)$ with each variable $v \in \V$ replaced by its value $\valOf{\ls}{v}$ in state $\ls$.

%	\begin{compactenum}
%	\item If $\locname$ has no handler over the action $\act$, then  $\hasAction{\ls}{\act}$ is simply $false$ (i.e., the handler does not exist regardless of the current interpretation).
%	
%   	\item  
%   	If $\locname$ has a handler $h$ over $\act$ with handler predicate $pred(d_1,\ldots,d_i)$, then
%   	$$\hasAction{\ls}{\act} \coloneqq pred(\valOf{s}{d_1},\ldots,\sigmaOf{d_i}).$$
%   	\editt{(or .... )}
%   	$$\hasAction{\ls}{\act} \coloneqq pred(\ls(d_1),\ldots,\ls(d_i)).$$
%   	   	\editt{(or .... )}
%   	   	$$\hasAction{\ls}{\act} \coloneqq \ls(pred).$$
%   	   	   	
%Note that, if $pred(d_1,\ldots,d_i)$ contained no uninterpreted functions, then $\hasAction{\ls}{\act}$ evaluates to $true$ or $false$ irrespective of the current interpretation. Alternatively, if it contained uninterpreted functions, then we are left with a boolean formula over functions in $\uFuns$ with domains drawn from $\sigma$.
%    
%	\end{compactenum}

%	\nour {Keeping this for now but it needs polishing.}

{\em Example.} Let $\uf(x,y)$ be an uninterpreted function over local \code{int} variables $x$ and $y$. Let the local state $\ls \coloneqq \{\locVar = \code{F}, x=1,y=2\}$, and let the local guard of action handler $\handler$ over action $\action$ in location \code{F} be 
$\guard \coloneqq \uf(x,y) > 7 \vee x=2$. Then $\hasAction{\ls}{\action}$ = \big((
$\uf(\valOf{\ls}{x},\valOf{\ls}{y}) > 7 \vee
\valOf{\ls}{x}=2)  = true)\big)$ 
which is 
$\big((\uf(1,2) > 7 \vee
1=2) = true\big)$ which simplifies to $\uf(1,2) > 7$.

%\editt{simplified the example}
%%%%%%%%%%%%%%%%%%%%%%%%%%%%%%%%%%%%%%%%%%%%%%%%%%%%%%%%%%%%%%%%%%%%%
%{\em Example.} Let $\uf_1(a,b,c)$ and $\uf_2(a,b)$ be uninterpreted functions over local \code{int} variables $a$, $b$ and $c$. Let $\ls \coloneqq \{\locVar = \code{Z}, a=1,b=2,c=3\}$, and let the local guard of action handler $h$ over action $\action$ in location \code{Z} be 
%$\guard \coloneqq \uf_1(a,b,c) \wedge (\uf_2(a,b) > 7 \vee a=2)$. Then $\hasAction{\ls}{\action}$ = (
%$\uf_1(\valOf{\ls}{a},\valOf{\ls}{b},\valOf{\ls}{c}) \wedge
%(\uf_2(\valOf{\ls}{a},\valOf{\ls}{b})  > 7 \vee
%\valOf{\ls}{a}=2)  = true)$ 
%which is 
%$(\uf_1(1,2,3) \wedge 
%(\uf_2(1,2) > 7 \vee
%1=2) = true)$.
%%%%%%%%%%%%%%%%%%%%%%%%%%%%%%%%%%%%%%%%%%%%%%%%%%%%%%%%%%%%%%%%%%%%%

%$\guard \coloneqq \code{\uf}_1\code{(a,b,c) \&\& (}\uf_2\code{(a,b) > 7 || a==2)}$. Then $\hasAction{\ls}{\act}$ is defined as follows: 
%$\code{\uf}_1\code{(\valOf{\ls}{a},\valOf{\ls}{b},\valOf{\ls}{c}) \&\& (}\uf_2\code{(\valOf{\ls}{a},\valOf{\ls}{b})  > 7|| \valOf{\ls}{a}==2)} $ 
%which is $ \code{\uf}_1\code{(1,2,3) \&\& (}\uf_2\code{(1,2) > 7|| 1==2)}$.

\item the predicate $\goesTo{\ls}{\action}{\ls'}$ indicating that $\ls$ goes to $\ls'$ on action $\action$.	The predicate $\goesTo{\ls}{\action}{\ls'}$ is defined as follows. 
Let $\update(\handler)$ denote the set of updates of the form $\lhs \coloneqq \rhs$ of handler $\handler$ over action $\action$. Then,	
	$\goesTo{\ls}{\action}{\ls'} \coloneqq \bigwedge_{\lhs \coloneqq \rhs \in \update(\handler)} \ls'(\lhs) = \rhs[\ls(\vars)/\vars].$\ourskip
	
%	\footnote{We treat \terma{goto} statements as an assignment to a special ``location'' variable.} 	
{\em Example.} Let the set of updates have the single update $x \coloneqq \uf(y,z)$ and $\ls,\ls'$ be $\{\locVar = \code{F}, x=1,y=2,z=3\}$ and $\{\locVar = \code{D},x=5,y=2,z=3\}$.
Then $\goesTo{\ls}{\action}{\ls'}$ is:
$\valOf{\ls'}{x} = \uf(\valOf{\ls}{y},\valOf{\ls}{z})$ which is 
$\uf(2,3)=5.$ \ourskip
\end{compactenum}
%\para{Manifestation Fingerprint of 
%\paraa{Encoding Disabled Local Transitions w.r.t. Interpretation $\interpretation$.}

\noindent An encoding of some disabled transition $t_{dis} = \trans{\ls}{\action}{\bot}$ in $\cex$
is defined as $ \existsTr{t_{dis}} = \reachable{\ls} \wedge \hasNoAction{\ls}{\action}$ where $\reachable{\ls}$ is as before
and the predicate $\hasNoAction{\ls}{\action}$, indicating that the process cannot perform action $\action$ from state $\ls$, is defined as follows:
$\hasNoAction{\ls}{\action} \coloneqq (\guard(\handler)[\valOf{\ls}{\vars}/\vars] = false).$
%\editt{double check this\footnote{\nour{revise this to be: collect ufs, make it uf.partialEval(src) = uf.eval(src)}}}
%where $\guard(\handler)[\valOf{\ls}{\vars}/\vars]$ is the guard $\guard(\handler)$ with each local variable $v \in \vars$ replaced by its value $\valOf{\ls}{v}$ in state $\ls$.\ourskip

The intuition behind breaking a transition's encoding to various predicates is that some \chwellbehavedness conditions %in \cite{quicksilver} 
leave parts of a transition unspecified. For instance, the predicate ``the local state $\ls$ can react to event $\event$'' corresponds to a local transition $\trans{\ls}{\reacting{\event}}{*} \in \itransitions$ with encoding $\reachable{\ls} \wedge \hasAction{\ls}{\reacting{\event}}$.\ourskip

Finally, to rule out any interpretation $\interpretation 
%\in \interpretations
$ that exhibits $\cex$, we add the constraint $\constraint = \neg \existsTr{\cex}$ to the \learner.\ourskip
%

%
%
%
%\subsection{Counter-Examples of Global Properties}
%Recall that $\pcomplete$ has the global semantics $\iglobalSemantics = (\iGQ,\gq_0,\events,\iGT, \iDisGT)$, where $\iGQ$ is the set of global states, $\gq_0$ is the initial global state, 
%$\events$ is the set of events,
%and  $\iGT$ is the set of global transitions.
%Further, recall that in the last two stages of the synthesis loop, we check if cutoff-sized system based on $\pcomplete$ is safe and live (i.e., $\isemantics \models \safetySpecc$ and $\isemantics \models \livenessSpec$). 
%
%
%Similar to counter-examples of local properties, a counter-example $\cex$ to safety (resp. liveness) is a ``subset'' of the global semantics $\iglobalSemantics$ such that $\cex \not \models \safetySpecc$ (resp. $\cex \not \models \livenessSpec$). We say that $\cex$ is a subset of $\iglobalSemantics$, denoted $\cex \subseteq \iglobalSemantics$, when it has a subset of its enabled and disabled global transitions, i.e., $\cex = (\iGQ,\gq_0,\events,\iGT' \subseteq \iGT,\iDisGTPrime \subseteq \iDisGT)$.
%
%
%
%
%\nour{remove livness from here}
%\mn{could possibly move to appendix?}
\para{Encoding \Counterexamples to Safety Properties}. Similar to the local semantics, we extend the definition of the global semantics $\iglobalSemantics{\pcomplete}{n}$ of a \aml system $\pcompletee{1} || \ldots || \pcompletee{n} || \process_e$
%consisting of $n$ identical processes $\process_{\interpretation,1}, \ldots,\process_{\interpretation,n}$ and an environment process $\process_e$
 to be $ \iglobalSemantics{\pcomplete}{n} =  (\iGQ,\gq_0,\events,\iGT, \iDisGT)$, where $\iGQ$, $\gq_0$, $\events$, and $\iGT$ are defined as before and $\iDisGT$ is the set of \emph{disabled global transitions} under the current interpretation $\interpretation$. Then,
%\Counterexamples to safety and liveness properties are formalized similar to these of \chwellbehavedness and \amenability, replacing the local semantics $\isemantics$ with $\iglobalSemantics$.
%\roopsha{Say that these counterexamples can be formalized smilarly, replacing local semantics with global semantics. Same for encoding. Skip everything below. }
%Recall that $\pcomplete$ has the global semantics $\iglobalSemantics = (\iGQ,\gq_0,\events,\iGT, \iDisGT)$, where $\iGQ$ is the set of global states, $\gq_0$ is the initial global state, 
%$\events$ is the set of events,
%and  $\iGT$ is the set of global transitions.
%Further, recall that in the last two stages of the synthesis loop, we check if cutoff-sized system based on $\pcomplete$ is safe and live (i.e., $\isemantics \models \safetySpecc$ and $\isemantics \models \livenessSpec$). 
%similar to \counterexamples of \chwellbehavedness and \amenability, 
a \counterexample $\cex$ to safety 
%(resp. liveness)
 is a ``subset'' of the \emph{global} semantics $\iglobalSemantics{\pcomplete}{c}$ such that $\cex \not \models \safetySpecc$. 
% (resp. $\cex \not \models \livenessSpec$). 
%We say that $\cex$ is a subset of $\iglobalSemantics$, denoted $\cex \subseteq \iglobalSemantics$, when it has a subset of its enabled and disabled global transitions, i.e., $\cex = (\iGQ,\gq_0,\events,\iGT' \subseteq \iGT,\iDisGTPrime \subseteq \iDisGT)$. 
Encoding of such a \counterexample $\cex$ is formalized as before, with the encoding of an enabled global transition $r$ in $\cex$
being a formula $\existsTr{\cex} \in \allConstraints$ computed as follows.
%that
%%over the atoms of the interpretation I that 
%(i) is satisfied under interpretation $\interpretation$ (i.e., $\interpretation \models \fconstraint$) and (ii) implies that $t \in \itransitions$. Informally, such formula $\fconstraint$ acts like an interpolant between the logical encoding of the interpretation $\interpretation$ and the logical encoding of the transition $t$. In what follows, for some transition $t =  \trans{\ls}{\action}{\ls'}$ based on action 
%
%
%
%\subsection{Encoding \Counterexamples of Global Properties}
%
%
%As before, we will focus on the encoding of a global transition, then lift that to the encoding of a \counterexample $\cex \subseteq \iglobalSemantics$.\ourskip
%
%\para{Encoding Global Transitions w.r.t. Interpretation $\mathbf{\interpretation}$.}
For some global transition $r = \trans{\gq}{\event}{\gq'}$, we denote by $\activeTrs(r)$ the local transitions that processes in $\gq$ locally use to end in $\gq'$. That is, $\activeTrs(r) = \{t \in \itransitions \mid \exists \process_{\interpretation,i} : t = \trans{\gq[i]}{\acting{\event}}{\gq'[i]} \vee t = \trans{\gq[i]}{\reacting{\event}}{\gq'[i]}\}$ %\editt{there must be a better way of doing this..}. 
We then define the encoding $\existsTr{r}$  as: %a constraint $\fconstraint \in \allConstraints$ as:
$
\existsTr{r} = \bigwedge_{t \in \activeTrs(r)}{\existsTr{t}}.
$

Note that the predicates $\reachable{\gq}$, $\hasAction{\gq}{\event}$, $\goesTo{\gq}{\event}{\gq'}$, and $\hasNoAction{\gq}{\event}$ as well as the encoding for the global disabled transitions can be defined similar to their counterparts discussed earlier.

\section{\Counterexample Extraction}
\label{sec:cinnabar}

%\begin{newtext}
%\nour{merge:}\para{\Counterexample Extraction and Encoding.} We equip each stage of the verification phase with an extraction strategy that generates a \emph{\counterexample} for when the candidate process fails at that stage. Informally, such \counterexamples are the ``relevant parts'' of the process model that caused some stage to fail. 
%\end{newtext}
Our tool specializes the synthesis procedure in
\algoref{synthAlgo} 
by using \kinarach 
\begin{wrapfigure}[14]{r}{0.45\textwidth}
	\vspace{-2em}
	\begin{minipage}{0.45\textwidth}
		\footnotesize
		\SetInd{0.28em}{0.43em}
		\begin{algorithm}[H]
			\SetKwFunction{synthisProc}{Extract}  
			\SetKwProg{myproc}{procedure}{}{}
			\myproc{\synthisProc{$\pcomplete, \prop$}}{
				
				$\prop'$ = $\text{\emph{makeDNF}}(\neg \prop$) \label{line:dnf}  
				
				%$\cex = \varnothing$, 
				$W = \varnothing$ \label{line:emptycex}
				
				\ForEach{$\cube \in cubes(\prop')$}{
					
					\If{$\isemantics{\pcomplete} \models \cube$\label{line:cubeOK}}{
						
						$\cubeWitness =\varnothing$ 
						
						\ForEach{$\literal \in literals(c)\label{line:startL}$}{
							
							$\literalWitness =\Call{witness}{\literal}$ \label{line:lw}
							
							$\cubeWitness = \cubeWitness \cup \{\literalWitness\}$  \label{line:endL}
						}
						$W = W \cup \{\cubeWitness\}$
					}
					$\cex = pickMinimal(W)$ \label{line:retMinimal}
					
					\Return $\cex$	
				}

			} % end of procedue
			\caption{
				\Counterexample Extraction.}
			\label{algo:extraction}
		\end{algorithm}
	\end{minipage}
\end{wrapfigure}
as the \teacher to check \chwellbehavedness, \amenability, and safety. 
For the remainder of this section, we will refer to \chwellbehavedness and \amenability conditions as \emph{local} properties and safety
 (and liveness) 
 specifications as \emph{global} properties.\ourskip

\para{Local Properties.} Given a local property $\prop$ expressed as first-order logic formulas over the local semantics of a \aml process, \tool extracts a counterexample $\cex$ according to \algoref{extraction}.

%\item Consider an arbitrary local property $\prop$.
First, we negate the property and express in disjunctive normal form (DNF): $\prop' = \neg \prop = \cube_1 \vee \cube_2 \vee \ldots$, 
  where each cube $\cube_i = \literal_1 \wedge \literal_2 \wedge \ldots$ is a conjunction of literals (\lineref{dnf}). 
  Then, for each cube $c$ satisfied under $\isemantics{\pcomplete}$ (\lineref{cubeOK}), extract a cube \emph{witness} $\cubeWitness$ that is a subset of the local semantics $\isemantics{\pcomplete}$ such that $\isemantics{\pcomplete} \models \cubeWitness$ (Lines \ref{line:startL} - \ref{line:endL}).
  This is done by extracting, for each literal $\literal$ in $\cube$, a minimal subset $\literalWitness$ of $\isemantics{\pcomplete}$ such that $\literalWitness \models \literal$ (\lineref{lw}). 
We say $\literalWitness$ is a \emph{minimal witness} of $\literal$ if any strict subset of $\literalWitness$ cannot be a witness for $\literal$ (i.e., $\forall \literalWitness' \subset \literalWitness: \literalWitness' \not \models \literal$). Finally
pick a minimal (in terms of size) cube witness  of some cube $\cube$ as a $\cex$ (\lineref{retMinimal}). Since $\cex \models \cube $ and $\cube \Rightarrow \neg \prop$, we know that $\cex \models \neg \prop$ (or equivalently, $\cex \not \models \prop$).

In this work, we carefully analyzed the \chwellbehavedness and cutoff amenability conditions and 
incorporated procedures to compute witnesses for their literals (i.e., the \code{witness} calls on \lineref{lw}). 
%\tool with the needed \counterexample extraction procedures for these literals.
We refer the interested reader to \appref{extractionApp} for complete details and illustrate one such counterexample extraction procedure using an example.

\noindent \emph{Example.} We present a simplified \chwellbehavedness condition and demonstrate the above procedure on it. 
Let the set of broadcast, partition, and consensus events be called the \emph{globally-synchronizing} events, denoted $\gsactions$.
Let $\pha(s)$ be the set of all ``phases'' containing local state $s$. The condition states that:
for each internal transition $\trans{s}{}{s’}$ that is accompanied by a
    reacting transition $\trans{s'}{R(\actb)}{s''}$ for some globally-synchronizing event $\actb$, and for each state $t$ in the same phase as $s$, state $t$ must have a reacting transition of event $\actb$. Formally:
\begin{multline*}
	\forall \actb \in \gsactions, s,s' \in S: \\
	\big{(} \trans{s}{}{s'} \in \T \wedge \trans{s'}{R(\actb)}{*} \in \T  \big{)} \Rightarrow 
	\big( \forall X \in \pha(s), 
	%:\ \firableIn{X}{\actb} \Rightarrow	\big{(} \forall 
	t \in X:\ \exists \trans{t}{R(\actb)}{*} \in \T
	%\big{)} 	
	\big).
	\end{multline*}
This condition is an example of a local property $\phi$ we want to extract \counterexamples for when it fails. The procedure is applied as follows:

\begin{compactenum}[ ]
\item[Step (1):] We first simplify $\phi$ to the following:
\begin{multline*}
\forall \actb \in \gsactions, s,s',t \in \LS, X \in \pha(s): \\ 
\big{(}
\trans{s}{}{s'} \in \T \wedge 
\trans{s'}{R(\actb)}{*} \in \T \wedge
%\trans{z}{A(\actb)}{*} \in \T \wedge
\samePhase{X}{s}{t} 
%\wedge \samePhase{X}{s}{z} 
\big{)}
\Rightarrow
\big{(}\exists \trans{t}{R(\actb)}{*} \in \T \big{)},
\end{multline*}
where $\samePhase{X}{s}{t}$ indicates that states $s$ and $t$ are in phase $X$ together. We then obtain the negation $\neg \phi$:
\begin{multline*}
\exists \actb \in \gsactions, s,s',t \in \LS, X \in \pha(s): \\ 
%\big{(}
\trans{s}{}{s'} \in \T \wedge 
\trans{s'}{R(\actb)}{*} \in \T \wedge
%\trans{z}{A(\actb)}{*} \in \T \wedge
\samePhase{X}{s}{t} 
%\wedge \samePhase{X}{s}{z} 
%\big{)}
\wedge 
\neg \exists \trans{t}{R(\actb)}{*} \in \T.
\end{multline*}
\item[Step (2):] The formula $\neg \phi$ is in DNF, and there is a cube for each instantiation of event $\actb \in \gsactions$, states $s,s',t \in \LS$, and phase $X$ that satisfies the formula $\neg \phi$. There are 4 literals. The literals ``$\trans{s}{}{s'} \in \T$ '' and    
``$\trans{s'}{R(\actb)}{*}\in \T$ '' can be witnessed by the corresponding transitions $\trans{s}{}{s'}$ and
$\trans{s'}{R(\actb)}{*}$, respectively. The literal
``$\neg \exists \trans{t}{R(\actb)}{*}\in \T$ '' can be witnessed by the \emph{disabled} transition 
$\trans{t}{R(\actb)}{\bot}$. 
%Witnesses for the literals $\samePhase{X}{s_a}{s_b}$ and $\neg \exists t \xrightsquigarrow{R(\actb)} *$  are defined as follows.
The witness for the literal $\samePhase{X}{s_a}{s_b}$ for some phase $X$ and local states $s_a$ and $s_b$ is more involved. It depends on the \emph{nature} of that phase. We analyzed the phase construction procedure given in \cite{Jaber.QuickSilver.OOSLA.2021} and distilled it as follows.
For each event $\event \in \gsactions$, we define its source (resp. destination) set to
%$\srcSet{\event} = \{\ls \mid \trans{\ls}{\acting{\event}}{\ls'} \in \T \vee 
%\trans{\ls}{\reacting{\event}}{\ls'}  \in \T \}$
be the set of states in $\LQ$ from (resp. to) which 
there exists a transition in $\T$ labeled with an acting or reacting action of event $\event$.
%Similarly, we define the destination set 
%$\destSet{\act} = \{\ls \mid \trans{\ls'}{\acting{\event}}{\ls} \in \T \vee 
%\trans{\ls'}{\reacting{\event}}{\ls}  \in \T \}$
%as the set of states in $\LQ$ to which a transition in $\T$ labeled with $\event$ exists.
Let $corePhases$ be the set of all source and destination sets of all globally-synchronizing actions. Then, two states $s_a$ and $s_b$ are in the same phase if:
\begin{compactenum}
\item they are part of some core phase, i.e., $\exists X \in corePhases: s_a,s_b \in X$, or,
\item they are in different core phases that are connected by an internal path, i.e., $\exists A,B \in corePhases: s_a, s_a' \in A \wedge s_b, s_b' \in B \wedge s_a' \rightsquigarrow s_b'$, where $ s_a' \rightsquigarrow s_b'$ is an internal path from $s_a'$ to $s_b'$.
\end{compactenum}
If $X$ is a core phase (i.e., case (A) holds), the \counterexample extraction procedure returns the phase itself. Otherwise, case (B) holds and the two core phases are recursively extracted as well as the internal path connecting them.

%The witness for the literal $\neg \exists t \xrightsquigarrow{R(\actb)} *$
%%, indicating that state $t$ cannot internally reach a state where 
%is extracted as follows. Let $P_{\text{int}}$ be the set of all the internal paths originating from $t$ and $S_{\text{int}}$ be the set of all states in these paths. Then, the witness is all paths in $P_{\text{int}}$, all \emph{disabled} internal transitions originating from $S_{\text{int}}$, and all \emph{disabled} reacting transitions on event $\actb$. Intuitively, this is the subset of the local semantics sufficient to show that $\neg \exists t \xrightsquigarrow{R(\actb)} *$  holds.

\item[Step (3)] The final step is to build a subset of the local semantics that include the extracted witnesses for all 4 literals.\ourskip
\end{compactenum}
\para{Global Properties.}
If a candidate process $\pcomplete$ meets its \chwellbehavedness and \amenability conditions, then it belongs to the \effdecidable fragment of \aml, and a cutoff $c$ exists. It then remains to check if the system
$\pcompletee{1} || \ldots || \pcompletee{n} || \process_e$
% \neww{$\pcomplete$ consisting of $\pcompletee{1},\ldots,\pcompletee{c}$ and environment process $\process_e$}
 is safe (i.e., $\iglobalSemantics{\pcomplete}{c} \models \safetySpecc$). %Below, we present safety and liveness properties. \ourskip

%\para{Safety Properties.} 
Safety properties $\safetySpec$ are specified by the system designer as (Boolean combinations of) \permissible safety specifications. Such properties are invariants that must hold in every reachable state in $\iglobalSemantics{\pcomplete}{c}$.
A \counterexample $\cex \subseteq \iglobalSemantics{\pcomplete}{c}$ to a safety property $\safetySpecc$ is a finite trace from the initial state $\gq_0$ to an error state $\gq_e$. %such that state $\gq_e$ violates $\safetySpec$. 
Such traces are extracted while constructing $\iglobalSemantics{\pcomplete}{c}$.%\footnote{We also check for \emph{deadlock-freedom}, i.e., that $\iglobalSemantics{\pcomplete}{c}$ does not contain any deadlock state where no event is enabled. A \counterexample to deadlock freedom is a trace to some deadlock state $\gq_d$ \emph{as well as} all disabled global transitions out $\gq_d$. After all, a deadlock state can be avoided if the system does not reach it, or can step out of it.} 

\section {Implementation and Evaluation}
\label{sec:evalAndImpl}
\subsection{Implementation}
\label{sec:implementation}
Our tool\footnote{\tool is publicly available  on Zenodo~\cite{zenodo}.} implements the architecture illustrated in \figref{flow}.
Additionally, it incorporates a liveness checker into the \teacher. Liveness properties $\livenessSpec$ ensure that the system makes progress and eventually reacts to various events. We refer the interested reader to \appref{liveness} for details on specifying liveness properties as well as extracting and encoding counterexamples to such properties.

%\begin{newtext}
%A liveness property is specified as a \buchi automaton \cite{buchi} or an LTL formula that accepts infinite traces in $\iglobalSemantics{\pcomplete}{c}$ violating the property.%\footnote{In fact, in \tool, the system designer can also specify liveness properties as (a restricted form of) Linear Time Logic (LTL) formulas that are then automatically translated to their \buchi automata equivalents.}
%
%As is the case with safety properties, if the system is not live (i.e., $\iglobalSemantics{\pcomplete}{c} \not \models \livenessSpec$), a \counterexample $\cex \subseteq \iglobalSemantics{\pcomplete}{c}$ is extracted and encoded. Such $\cex$ is a ``lasso-shaped'' subset of the global semantics that represents an infinite, \fair, accepting trace violating $\livenessSpec$. An infinite trace is \fair if it represents a realistic infinite execution of the system as opposed to one that is merely an artifact of the non-determinism when picking which system event should be executed next. Such $\cex$ consists of two parts: a fair, accepting \emph{cycle} in $\iglobalSemantics{\pcomplete}{c}$, and a finite \emph{stem} from the initial state $\gq_0$ to a state in that cycle.
%Identifying fair, accepting cycles and their stems is a standard technique in the literature and we defer its details to \appref{liveness}.
%Note that $\cex$ \emph{additionally} contains all disabled global transitions from every state in the cycle. Intuitively, enabling any such disabled transition renders the infinite trace unfair, and hence rules out the liveness violation.
%\end{newtext}

\subsection{Evaluation}
\label{sec:evaluation}
%\para{\tool}. We implement our ideas in \tool, using \kinarach as the \teacher. From the set of 163,840,000 possible completions of the Distributed Store sketch in \figref{distStore}, \tool is able to find a correct one in 2.2 minutes and 203 iterations. In contrast, a naive exploration of all possible completions times out after 15 minutes. This demonstrates the impact of \counterexample extraction and encoding on effectively pruning the search space.\mn{removed para about nature of violations.} 

%\subsection{Evaluation}

%In this section, we first describe the set of benchmarks that our evaluation is based on, and our experimental setup. Then, we investigate the performance of \tool through the following research questions:

In this section, we investigate \tool's performance. 
%We start by a brief description of our benchmarks and the experimental setup. Then, w
We study the impact of \tool's \counterexample extraction and encoding, as well as the choice of uninterpreted functions, on performance. Finally, we examine how \tool's iterations are distributed across the different types of \counterexamples. %Finally, we investigate the performance variation w.r.t. the choice of uninterpreted functions.
\ourskip

%and the value of its \counterexample extraction and uniform encoding. We start by posing our research questions, presenting the set of agreement-based systems we perform synthesis for, 
%the \emph{enumeration-based} baseline we compare against, 
%as well as our experimental step.\ourskip
%\para{Research Questions.}
%The research questions we want to tackle in this empirical evaluation are:
%\begin{description}
%	\item[RQ1] What is the impact of \tool's \counterexample extraction and encoding on its overall performance compared to a baseline of excluding just the incorrect candidate solution?
%	
%%	\item[RQ2] What percentage of \tool's effort goes into finding a candidate process in the \effdecidable?
%	\item[RQ2] How are \tool's iterations distributed across the different types of \counterexamples?
%%	
%%	\item[RQ3] What impact does the choice of unspecified expressions have on \tool's performance?\ourskip
%	\item[RQ3] How does \tool's performance vary w.r.t. the choice of unspecified expressions?\ourskip
%\end{description}

\para{Benchmarks.} The benchmarks we use are process sketches based on the benchmarks presented in
%We use the \aml systems presented in
\cite{Jaber.QuickSilver.OOSLA.2021}. 
%Below, be briefly describe the each benchmark, and
We refer the reader \appref{eval} for a description of each benchmark's functionality, its safety and liveness specifications, and the unspecified functionality in the sketch. An example \aml sketch and its completion in also included in \appref{eval}.
\ourskip

\para{Experimental Setup.} 
%Each of our benchmarks includes 6 to 10 ``candidate expressions'', corresponding to expressions that a designer might reasonably leave unspecified. Each experimental configuration then chooses a subset of these candidate expressions to leave as {\em real} unspecified expressions for \tool to complete. Answering RQ1 requires understanding how efficiently \tool can synthesize correct completions. 
To ensure that our reported results are not dependent on a particular choice of uninterpreted functions, we create a set of \emph{variants} for each benchmark as follows. For each benchmark, we first pick a set $ue$ of ``candidate uninterpreted functions'', corresponding to expressions that a designer might reasonably leave unspecified.
%Let $ue$ denote the set of all candidate holes in a benchmark and $\powerset(ue)$ denote the set of all non-empty subsets of $ue$. 
Then, for each subset $e$ in the set $\powerset(ue)$ of all non-empty subsets of $ue$, we create a variant of the benchmark where the uninterpreted functions in $e$ are included in the sketch.
%
%\para{Experimental Setup.} Each of our benchmarks includes 6 to 10 ``candidate expressions'', corresponding to expressions that a designer might reasonably leave unspecified. Each experimental configuration then chooses a subset of these candidate expressions to leave as {\em real} unspecified expressions for \tool to complete. Answering RQ1 requires understanding how efficiently \tool can synthesize correct completions. 
%To ensure that our reported results are not dependent on the choice of such unspecified expressions, we create a set of \emph{variants} for each benchmark as follows. Let $ue$ denote the set of all candidate holes in a benchmark and $\powerset(ue)$ denote the set of all non-empty subsets of $ue$. Then, for each subset $e \in \powerset(ue)$, we create a variant of the benchmark where the expressions in $e$ are left unspecified.
%
%and $m$ be the cardinality of $ue$. Then, for each $k \in [1..m]$ and for each subset we create a variant for that where 
%we generate all variations of all $k$-combinations of these expressions for all $k \in [1..m]$ where $m$ is the maximum number of unspecified expressions in that benchmark. 
%\neww{We run each variant five times and report the median of these runs}. 
%Additionally, 
We set a timeout of 15 minutes when running any variant and %Finally,  we 
conduct our experiments on a MacBook Pro with 2 GHz Quad-Core Intel Core i5 and 16 GB of RAM.\ourskip
\begin{figure}[t]
%	\squeezecaption
%	\squeezecaption
%	\squeezecaption
%	\squeezecaption
%	\squeezecaption
	
\hspace{1em}
% trim=left botm right top.
%\fbox{
\includegraphics[width=0.85\linewidth, clip, trim=0.3cm 1cm 0.3cm 1.3cm]{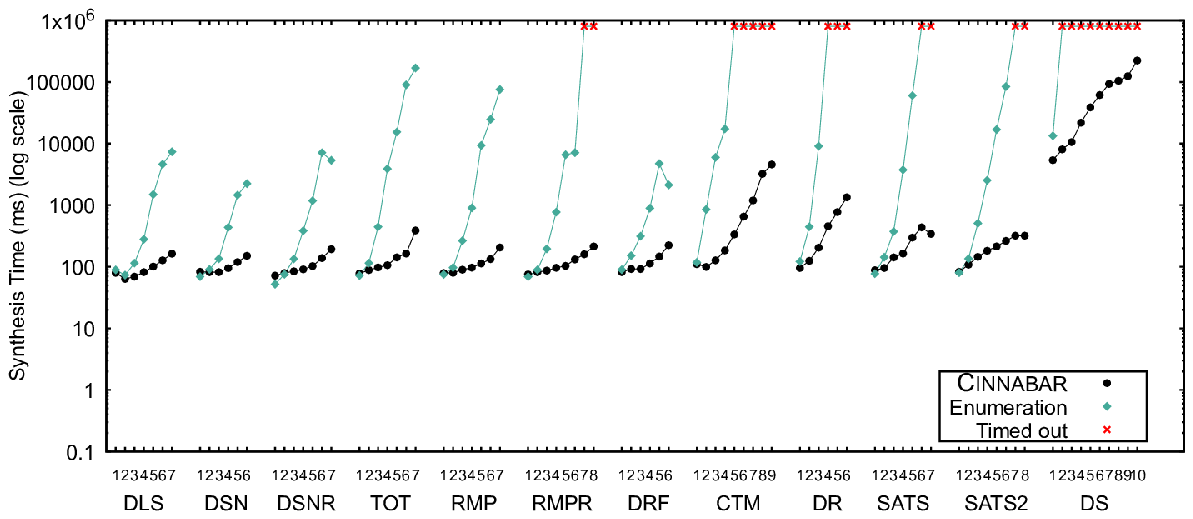}
%}
\caption{\tool's performance compared to enumeration-based synthesis. The systems studied are: Distributed Store (DS),
Consortium (CTM),
Distributed Lock Service (DLS),
Distributed Register (DR),
Two-Object Tracker (TOT),
Distributed Robot Flocking (DRF),
variants Small Aircraft Transportation System Landing Protocol (SATS, SATS2),
variants of Distributed Sensor Network (DSN, DSNR), and variants
of Robotics Motion Planner (RMP, RMPR).
For each benchmark, the $i$-th point denotes the average runtime for all variants with $i$ uninterpreted functions.}
\label{fig:comparison}

%\squeezecaption
%\squeezecaption
%\squeezecaption
%\squeezecaption
%\squeezecaption
%\squeezecaption
%\squeezecaption
%\squeezecaption
%\squeezecaption
%\squeezecaption
%\squeezecaption
\end{figure}

%\para{Enumeration-Based Synthesis.} 
\para{Effect of \Counterexample Extraction and Encoding.} As our baseline, we consider a synthesis loop where the \learner enumerates interpretations until a correct interpretation is found. %process. 
If some interpreted process sketch $\pcomplete$ fails a property at any stage, we add the constraint $\constraint = \neg \interpretation$ to the \learner. This effectively eliminates one interpretation at a time, as opposed to %\changee{one \counterexample}{
all interpretations that exhibit the given \counterexample
%} 
at a time (as done by our encoder). In \figref{comparison}, we present a comparison of \tool's runtime compared to this enumeration-based baseline.
%\para{Value of Extracting and Encoding \Counterexamples.} To answer RQ1, we evaluate how an encoding-aware \tool performs compared to an enumeration-based synthesis loop, as shown in \figref{comparison}. 
We make the following observations.
 While the runtimes of both enumeration-based synthesis and \tool grow exponentially when increasing the number of uninterpreted functions, \tool outperforms enumeration-based synthesis in almost all scenarios. Only for variants with a single uninterpreted function we observed cases where enumeration-based synthesis found a correct solution faster than \tool 
(e.g., as in DSNR with one uninterpreted function). This is due to the additional time spent extracting and encoding \counterexamples. 
However, the value of the \counterexample extraction and encoding becomes clearly apparent with larger number of unspecified expressions as the number of interpretations grows much larger and it becomes infeasible to just enumerate them.
 %While the runtimes of both enumeration-based synthesis and \tool grow exponentially when increasing the number of unspecified expressions, the latter scales more gracefully and is able to 
 Furthermore, \tool is able to perform synthesis for any variant of our benchmarks in under 9 minutes.\ourskip

\begin{figure}[t]
%\squeezecaption
\hspace{1em}
% trim=left botm right top.
%\fbox{
\includegraphics[width=0.85\linewidth, clip, trim=0.3cm 1cm 0.3cm 1.3cm]{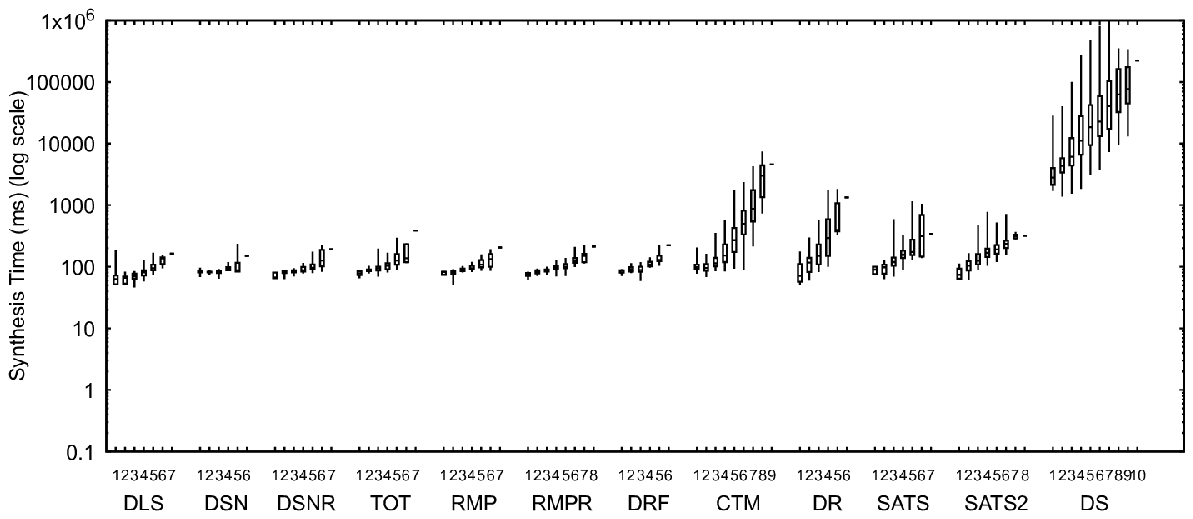}
%}
\caption{Effect of the choice of uninterpreted functions on synthesis time. For some benchmark and some number $m$ of uninterpreted functions, the $m$-th box-and-whiskers plot presents, from bottom to top, the minimum, first quartile, median, third quartile, and maximum synthesis run time across the run times of all variants of that benchmark with $m$ uninterpreted functions.}

%
%
%
%\todo{figure out a caption} write this carefully to signify that this is about the runtime of variants, not multiple runs of the same variant}
\label{fig:timeVariation}

%\squeezecaption
%\squeezecaption
%\squeezecaption
%\squeezecaption
%\squeezecaption
\end{figure}

\para{Effect of the Choice of Uninterpreted Functions.} 
In \figref{timeVariation}, for each benchmark, 
we examine the variation of synthesis runtime across variants with the same number of uninterpreted functions. 
As shown in the figure, in some cases (e.g., CTM and DS), the variation is more noticeable. The main factor contributing to this is that uninterpreted functions present different overhead on synthesis based on their nature. For instance, an uninterpreted function corresponding to a \termb{lhs} of some assignment expression is more expensive to synthesize compared to an uninterpreted function corresponding to a target of some \terma{goto} statement, as the latter has a smaller search space.\ourskip

\begin{figure}[t]
%	\squeezecaption
%	\squeezecaption
%	\squeezecaption
%	\squeezecaption
%	\squeezecaption
%	\squeezecaption
%	\squeezecaption
\centering
% trim=left botm right top.
%\fbox{
%\includegraphics[width=0.85\linewidth, clip, trim=0cm 0cm 0cm 2cm]{heatmapv2.pdf}
\includegraphics[width=0.85\linewidth]{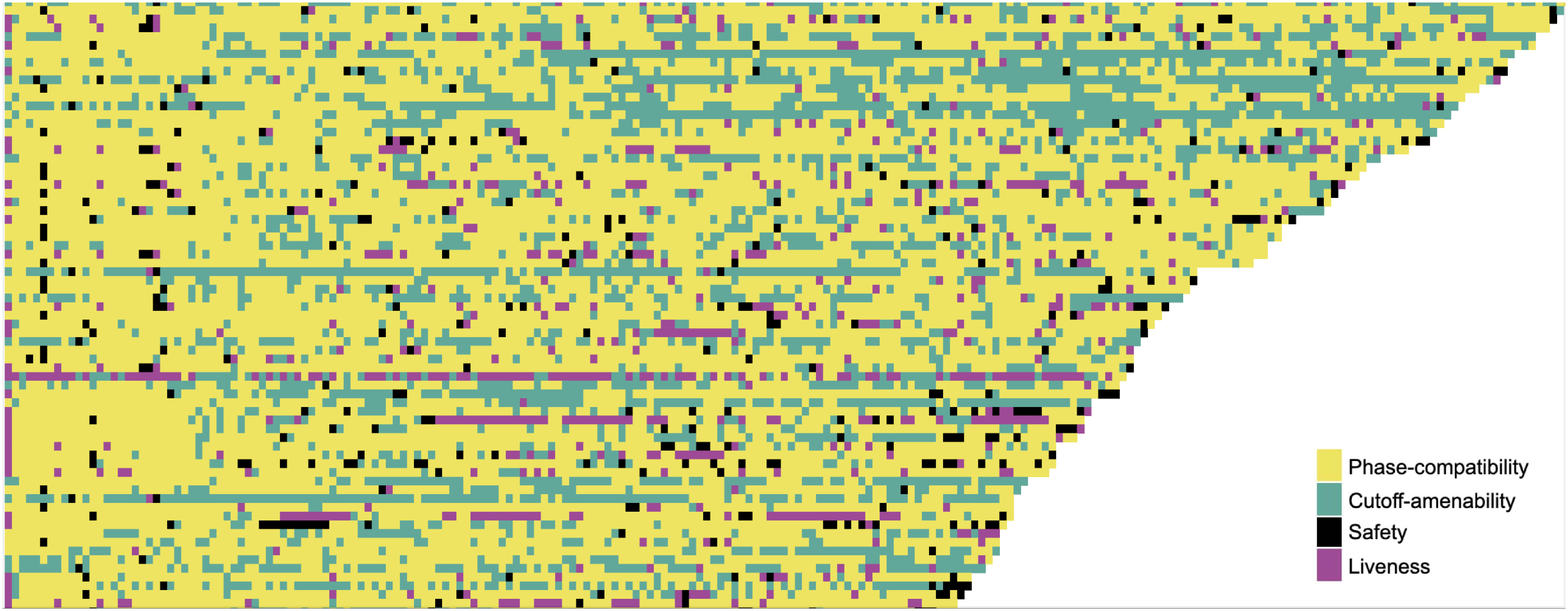}
%}
\caption{A property-based visualization of \tool's iterations
 %when executing 
 for a representative subset of the variants. %\nour{representative snippet?} 
%A snippet of the violations distribution throughout the execution of \tool.
Each line corresponds a \tool's execution of a synthesis variant of a benchmark. From left to right, each line starts with iteration 1, ends  with the iteration where a correct interpretation was found, and is colored to indicate nature of violations encountered throughout the execution. 
For instance, the line \includegraphics[height=\myMheight]{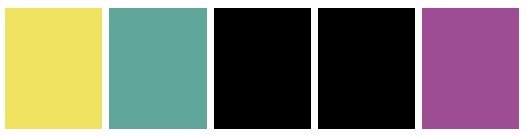} would indicate that \tool encountered a \chwellbehavedness violation in iteration 1, then a \amenability in iteration 2, ...,  and finally was able to find a correct interpretation in iteration 6.} 
%Black: 0, 0, 0 (#000000)
%Purple: 170, 68, 153 (#AA4499)
%Teal: 68, 170, 153 (#44AA99)
%Yellow: 240, 228, 66 (#F0E442)
\label{fig:heatmap}
%\squeezecaption
%\squeezecaption
%\squeezecaption

\end{figure}

\para{\Counterexample Distribution on Iterations.}  In \figref{heatmap}, we illustrate the different types of \counterexamples encountered throughout \tool's iterations.
%We now focus on the nature of violations encountered throughout a run of \tool with \counterexample extraction and encoding. The goal here is to answer  RQ2 by observing how \tool converges to a correct solution. 
We make the following observations. First, \tool spends most of its iterations ruling out \chwellbehavedness violations. This is expected as checking \chwellbehavedness is the first stage in our synthesis loop. Since a \chwellbehaved system moves in a structured way between its phases, this stage rules out all arbitrary completions that prohibit processes from advancing through the phases.
%\item The second most-common violation is \amenability. Unlike \chwellbehavedness, \amenability depends on the safety specification---in particular, on the properties of paths from initial states to the error states determined by the safety specification.
%\swen{rather than the (obvious) next sentence, maybe say something like ``In contrast to phase-compatibility, this property depends on the safety specification, in particular on syntactic properties of paths from initial states to error states.''? (And then maybe the explanation in the next point can be adjusted/simplified based on that)} Once a candidate process is deemed \chwellbehaved, it must additionally be \amenable.
Furthermore, there are fewer safety violations than any other type of violations. Once an interpreted process sketch is in the \effdecidable fragment of \aml, it is more likely to be safe. There are two factors that contribute to this: (i) \chwellbehaved systems move in a structured way and are more likely to be ``closer'' to a correct version of the system, and (ii) because \amenability depends on the safety specification, satisfying \amenability means the interpreted process sketch is more likely to be correct with respect to the safety property already.
%the \amenability conditions are designed for a specific class of safety properties, hence, if a specific \amenability condition holds for a candidate process,
%%if a cutoff is computable, 
%then the candidate process is more likely to be correct w.r.t. the corresponding class of safety properties.
Finally, eliminating liveness violations ensures that \tool is able to synthesize higher-quality completions. As shown in the figure, liveness violations are often encountered in the very first iteration, as the SMT-based \learner tends to favor interpretations with disabled guards that trivially satisfy \chwellbehavedness, \amenability, and safety properties.\ourskip

\para{Usability.} \mn{addressing the ``what happens when synthesis fails} If \tool fails to synthesize a correct completion, the designer can \emph{replace} existing expressions in the sketch with uninterpreted functions, allowing \tool to explore a larger set of possible candidate completions. 

Finally, while the supported uninterpreted functions may not correspond to large segments of the code or complex control-flow constructs, they are the main ``knobs'' that the designer needs to turn to ensure that their systems belong to the \effdecidable fragment of \aml. 

\section{Related Work}
\label{sec:related}

%Our technique uses program synthesis to aid the system designer in ensuring that their \aml models belong to the \effdecidable fragment of \aml. Hence,
%\mn{can remove this para for length.}We look at related work through three lenses: (i) approaches that aid designers through decidable verification, (ii) approaches that target parameterized synthesis, and (iii) approaches that target synthesis for distributed systems in general.\ourskip

\para{Aiding System Designers via Decidable Verification.} Ivy~\cite{Padon.IvySafetyVerification.PLDI.2016} adopts an interactive approach to aid the designer in searching for inductive invariants for their systems. 
Ivy translates the system model and its invariant to EPR~\cite{EPR}, and looks for a \emph{counterexample-to-induction} (CTI). The designer adjusts the invariant to eliminate that CTI and Ivy starts over. 
%Ivy restricts its modeling language and invariants to ensure that they can be translated to a decidable fragment of first-order logic (namely, effectively propositional logic (EPR)~\cite{EPR}).
%The designer provides a model and an initial invariant that are translated to EPR and checked automatically for correctness. If the check fails, then the invariant is not inductive and the designer is presented with a \emph{counterexample-to-induction}. The designer then adjusts the invariant and starts over. 
I4 \cite{I4} builds on Ivy by first considering a fixed system size, automatically generating a potential inductive invariant, and using Ivy to check if that invariant is also valid for any system size. 
The approach in \cite{damian.communicationclosed.CAV.2019} identifies a class of  asynchronous systems that can be reduced to an equivalent synchronized system modeled in the Heard-Of Model \cite{HOmodel}. The designer manually annotates the asynchronous system to facilitate the reduction, and encodes the resulting Heard-Of model in the $\mathbb{CL}$ \cite{Druagoi.LogicbasedFrameworkVerifying.X.2014} logic which has a semi-decision procedure.
These approaches differ from ours in two ways. First, the designer needs to manually provide/manipulate inductive invariants and/or annotations to eventually enable decidable verification. Second, these approaches are ``verification only'': they require a \emph{fully-specified} model that either meets or violates its correctness properties and the designer is responsible for adjusting the model if verification fails. \tool, on the other hand, accepts a sketch that is then completed to meet its properties. \ourskip

\para{Parameterized Synthesis.} 
%The parameterized synthesis problem can be approached in one of two ways: reducing it to the fixed-size synthesis problem via cutoff results, or developing decision procedures for it directly. 
Jacobs and Bloem  \cite{Jacobs.ParameterizedSynthesis.X.2012} introduced a general approach for parameterized synthesis based on cutoffs, where they use an underlying fixed-size synthesis procedure that is required to guarantee that the conditions for cutoffs are met by the synthesized implementation. 
Our approach can be seen as an instantiation of this approach, as one of the stages in our multi-stage counterexample-based loop 
%is responsible for ensuring
 ensures that \amenability conditions hold on any candidate process. Other approaches that tackle the parameterized synthesis problem without cutoff results are more specialized. 
For instance, the approach in \cite{Lazic.SynthesisDistributedAlgorithms.X.2018} adopts a CEGIS-based synthesis strategy where the designer provides a threshold automaton with some parameters unspecified. Synthesis completes the model and uses the parameterized model checker in \cite{Konnov.ShortCounterexampleProperty.X.2017} to check the system. 
A similar idea, but based on the notion of well-structured transition systems, is used for the automatic \emph{repair} of parameterized systems in~\cite{jacobs2022automatic}.
The approach in \cite{Klinkhamer.Synthesizing.Parameterized.FSEN.2017} targets parameterized synthesis for self-stabilizing rings, and shows that the problem is decidable even when the corresponding parameterized verification problem is not. The designer provides a set of legitimate states and the size of the template process, and the procedure yields a completed self-stabilizing template. A similar approach for more general topologies is presented in~\cite{MirzaieFJB20}.
Bertrand et al. \cite{bertrand2018controlling} target systems composed of an unbounded number of agents that are fully specified and one underspecified controller process. The synthesis goal is to synthesize a controller that controls all agents uniformly and guides them to a specific desired state. Markgraf et al. \cite{Markgraf.Parameterized.Synthesis.Safety.Properties.2020} also target synthesis of controllers by posing the problem as an infinite-duration 2-player game and utilize regular model checking and the L* algorithm \cite{LStarAlgo} to learn correct-by-design controllers.
%correct by construction
%infinite-duration 2 player game
%In this paper, we are interested in automatically synthesizing correct parameterized systems with a safety guarantee. In this setting, parameterized systems
%are only partially specified, and the task of a synthesis algorithm is to “fill in” the
%missing specification in such a way that the desired property is satisfied. Synthesis algorithms aim to produce a correct-by-construction implementation of some
%formal properties in a fully automatic fashion, thereby saving the need for performing a further verification step
These approaches are not applicable to our setup as they do not admit distributed agreement-based systems (modeled in \aml). \ourskip

\para{Synthesis of Distributed Systems with a Fixed Number of Processes.}
Various approaches focus on automated synthesis of distributed
systems with a \emph{fixed} number of processes~\cite{Alur.AutomaticSynthesisDistributed.X.2017,Alur.AutomaticCompletionDistributed.X.2015,Alur.SynthesizingFinitestateProtocols.X.2014,Damm.AutomaticCompositionalSynthesis.X.2014,Udupa.TRANSITSpecifyingProtocols.X.2013}. While such approaches deploy a similar \counterexample-guided strategy to complete a user-provided sketch, they do not provide parameterized correctness guarantees nor the necessary agreement primitives needed to model distributed agreement-based systems.
\bibliographystyle{splncs04}
\bibliography{main}
\newpage
\appendix
%\section{Details for Counter-Example Extraction}
\section{Extracting \Counterexamples to \CHwellbehavedness and \AMenability Conditions}
\label{app:extractionApp}
In this section, we first recap the \chwellbehavedness %from \cite{Jaber.QuickSilver.OOSLA.2021}
, and discuss their \counterexample extraction and encoding functions. We then give aa similar statement to the \amenability conditions.
\subsection{\CHwellbehavedness Conditions}

Let the set of broadcast, partition, and consensus events be called the \emph{globally-synchronizing} events, denoted $\gsactions$.
Additionally, For each event $\event \in \gsactions$, we define its source set $\srcSet{\event} = \{\ls \mid \trans{\ls}{\acting{\event}}{\ls'} \in \T \vee 
\trans{\ls}{\reacting{\event}}{\ls'}  \in \T \}$
be the set of states in $\LQ$ from which 
there exists a transition in $\T$ labeled with $\event$.
Similarly, we define the destination set 
$\destSet{\act} = \{\ls \mid \trans{\ls'}{\acting{\event}}{\ls} \in \T \vee 
\trans{\ls'}{\reacting{\event}}{\ls}  \in \T \}$
as the set of states in $\LQ$ to which a transition in $\T$ labeled with $\event$ exists.
Furthermore, for some event $\event$ and some subset $X$ of the local state space $\LS$, we say that $\event$ is {\em \firable} in $X$ if some state in $X$ has an acting transition of $\event$. We will denote this as $\firableIn{X}{\event}$. Finally, let $\ls \xrightsquigarrow{\reacting{\event}} {\ls'}$ denote a path $\ls \rightarrow^* \ls' \xrightarrow{\reacting{\event}} \ls''$ such that  $\ls \rightarrow^* \ls'$ is a path of internal transitions and $\trans{\ls'}{\reacting{\event}}{\ls''}$. 

\begin{definition}[\CHwellbehavedness Conditions \cite{Jaber.QuickSilver.OOSLA.2021}]

\begin{compactenum}[(1)]
	\item[] %force the items to start on a new line!

	\item Every state $\ls \in \LS$ which has an acting transition $\trans{\ls}{\acting{\event}}{\ls’}$  must also have a 
	 corresponding reacting transition $\trans{\ls}{\reacting{\event}}{\ls’'}$:	
	$$\forall \event \in \gsactions: \trans{\ls}{\acting{\event}}{\ls'} \implies \trans{\ls}{\reacting{\event}}{\ls''}$$
  \item For each acting transition $\trans{\ls}{\acting{\act}}{\ls'}$ that is accompanied by a
  	%n outgoing 
  	reacting transition $\trans{\ls'}{\reacting{\actb}}{\ls''}$ such that $\actb$ is \firable in the set %$\{v \mid \trans{v'}{A(\act)}{v} \in \T \ \vee \
  	%\trans{v'}{R(\act)}{v} \in \T\}$ 
  	$\destSet{\act}$ of {\em destination states} of event $\act$,   
  %includes some state $x$ with an acting transition $\trans{x}{A(\actb)}{x'}$
 (i) if there are other acting transitions $\trans{t}{\acting{\act}}{t'}$ for event $\act$, all of them must transition to a state $t'$ with a
 %n outgoing 
 reacting transition $\trans{t'}{\reacting{\actb}}{t''}$ of event $\actb$ and 
(ii) for every reacting transition $ \trans{u}{R(\act)}{u'}$ of $\act$, there must be a path from $u'$ to a state with a
%n outgoing 
reacting transition of event $\actb$.
\begin{multline*}
\; \; \; \forall \act,\actb \in \gsactions. 
\Big{(} \trans{s}{A(\act)}{s'} \wedge  s'  \xrightarrow{R(\actb)} s'' \Big{)} \implies \\ 
\firableIn{\destSet{\act}}{\actb} \implies
%\neww{ \Big{(} \exists x \in dst_\act. \trans{x}{A(\actb)}{x'} \Big{)} \implies }
%\Big{(} dst_\act \cap snd_\actb \neq \varnothing \Big{)} \implies \\ 
\Big{(} \forall \trans{t}{A(\act)}{t'}. \exists \trans{t'}{R(\actb)}{t''} \Big{)} \wedge \Big{(} \forall \trans{u}{R(\act)}{u'}. \exists u' \xrightsquigarrow{R(\actb)} u'' \Big{)} 
\end{multline*}
    \item For each internal transition $\trans{s}{}{s’}$ that is accompanied by a
    reacting transition $\trans{s'}{R(\actb)}{s''}$ and for each state $t$ in the same phase as $s$, if event $\actb$ is \firable in that phase, then $t$ must have a path to a state with a reacting transition of event $\actb$. Let $\pha(s)$ be the set of all phases containing $s$:
\begin{multline*}
	\forall \actb \in \gsactions: \Big{(} \trans{s}{}{s'} \wedge \trans{s'}{R(\actb)}{s''} \Big{)} \implies \\
	\forall X \in \pha(s).\ \firableIn{X}{\actb} \implies 	\Big{(} \forall t \in X.\ \exists t \xrightsquigarrow{R(\actb)} t' \Big{)}
	\end{multline*}
\end{compactenum}
\label{def:wbcpaper}
\end{definition}

\subsection{Extracting Witnesses in the Negation of \CHwellbehavedness Conditions}
We obtain the negation of the conditions in order to define the literals used. 

\subsubsection{Witnessing a Violation of Condition (1)}
The negation of this condition is simply: 
	$$\exists \event \in \gsactions: \trans{\ls}{\acting{\event}}{\ls'} \in \T \wedge \trans{\ls}{\reacting{\event}}{\ls''} \not \in \T$$
Since this formula is essentially in DNF, the witness is simply a state $\ls$ which has the acting transition on some event $\event$ but not the reacting one.

\subsubsection{Witnessing a Violation of Condition (2)}
We first obtain the equivalent condition (ref. \lemref{prenexform2}):
\begin{multline*}
\forall \act,\actb \in \gsactions, s,s',t,z,z' \in \LS: \\ 
(\trans{s}{A(\act)}{s'} \wedge
\trans{s'}{R(\actb)} {*} \wedge
\trans{z}{A/R(\act)}{z'} \wedge 
\trans{z'}{A(\actb)}{*} \wedge 
\trans{t}{R(\act)}{t'}) 
\implies 
\exists t' \xrightsquigarrow{R(\actb)} *
\end{multline*}
We focus in the ``reacting'' part of the condition. The other part is straightforward. The negation of the condition is: 
\begin{multline*}
\exists \act,\actb \in \gsactions, s,s',t,z,z' \in \LS: \\ 
\trans{s}{A(\act)}{s'} \wedge
\trans{s'}{R(\actb)} {*} \wedge
\trans{z}{A/R(\act)}{z'} \wedge 
\trans{z'}{A(\actb)}{*} \wedge 
\trans{t}{R(\act)}{t'}
\wedge
\neg \exists t' \xrightsquigarrow{R(\actb)} *
\end{multline*}
And, as before, the witness is any two events $\act,\actb \in \gsactions$ and states $s,s',t,z,z' \in \LS$ that satisfy the formula. Note that the witness for $\neg \exists t' \xrightsquigarrow{R(\actb)} *$ is extracted as follows. Let $P_{int}$ be all the internal paths originating from $t'$ and $S_{int}$ be the set of all states in these paths. Then, the witness is all paths in $P_{int}$, all \emph{disabled} internal transitions originating from $S_{int}$, and all \emph{disabled} reacting transitions on event $\actb$. Intuitively, this is the subset of the local semantics for which $\neg \exists t' \xrightsquigarrow{R(\actb)} *$  holds.

\subsubsection{Witnessing a Violation of Condition (3)}
We again first obtain a simpler, equivalent formula (ref. \lemref{prenexform3}):
\begin{multline*}
\forall \actb \in \gsactions, s,s',t,z \in \LS, X \in \pha(s): \\ 
\big{(}
\trans{s}{}{s'} \wedge 
\trans{s'}{R(\actb)}{*} \wedge
\trans{z}{A(\actb)}{*} \wedge
\samePhase{X}{s}{t} \wedge 
\samePhase{X}{s}{z} 
\big{)}
\Rightarrow
\big{(}\exists t \xrightsquigarrow{R(\actb)} * \big{)}
\end{multline*}
where $\samePhase{X}{s}{t}$ indicates that states $s$ and $t$ are in some phase $X$ together. The negation of the condition is as before, and the interesting part is extracting a witness for $\samePhase{X}{s_a}{s_b}$ for some phase $X$ and local states $s_a$ and $s_b$. This is done as follows. Let $corePhases$ be the set of all source and destination sets of all 
globally-synchronizing actions. Two states $s_a$ and $s_b$ are in the same phase if:

\begin{compactenum}
\item they are part of some core phase: $\exists X \in corePhases: s_a,s_b \in X$, or,
\item they are in different core phases that are connected by an internal path: $\exists A,B \in corePhases: s_a, s_a' \in A \wedge s_b, s_b' \in B \wedge s_a' \rightsquigarrow s_b'$.
\end{compactenum}

In case (A), the \counterexample extraction procedure simply returns the core phase itself and in case (B) the two core phases are returned as well as the internal path between them.

\subsection{\AMenability Conditions}
Jaber et. al \cite{Jaber.QuickSilver.OOSLA.2021} define a notion of {\em independence} of local transitions. Informally, \freech transitions are ones a process can take without requiring the \emph{existence} of other processes in certain states (e.g., sending a broadcast). A path is independent if it consists of independent transitions.

\begin{definition}[\AMenability Conditions \cite{Jaber.QuickSilver.OOSLA.2021}]
Let $\process$ be a \chwellbehaved process, $\safetySpec$ a \permissible specification, and $\mathcal{F}$ the set of \freech simple paths from $\ls_0$ to a state $\ls \in \states(f)$%$P$%\roopsha{Should ideally have a notation for semantics of $P$}
. We require either of the following to hold.\ourskip

\begin{compactenum}
\item All paths from $\ls_0$ to $\states(f)$ are \freech, or, 
\item For every transition $\trans{\ls_s}{}{\ls_d}$ such that  $\ls_s \neq \ls_d$ and $\ls_s$ is a state in some path $p \in \mathcal{F}$, either 
    \begin{compactenum}[(a)]
    \item the state $\ls_d$ is in $p$ and the transition $\trans{\ls_s}{}{\ls_d}$ is \freech, or,
    \item the state $\ls_d$ is not in $p$ and all paths out of $\ls_d$ lead back to $\ls_s$ via \freech transitions.
  \end{compactenum}
\end{compactenum}
\label{def:amenabilitypaper}
\end{definition}

\subsection{Extracting Witnesses in the Negation of \Amenability Conditions}

The negation of the above condition is:
\begin{compactenum}
\item [($x$)] some path from $\ls_0$ to $\states(f)$ is not \freech, and, 
\item [($y$)] some transition $\trans{\ls_s}{}{\ls_d}$ such that  $\ls_s \neq \ls_d$ and $\ls_s$ is a state in some path $p \in \mathcal{F}$ and
    \begin{compactenum}[(a)]
    \item the state $\ls_d$ is in $p$ and the transition $\trans{\ls_s}{}{\ls_d}$ is not \freech, or ,
    \item the state $\ls_d$ is not in $p$ and some path out of $\ls_d$ does not lead back to $\ls_s$, or leads back to $\ls_s$ but has a non-\freech transition.
  \end{compactenum}
\end{compactenum}

Interestingly, the condition can fail in two ways: either $x \wedge y \wedge a$ or $x \wedge y \wedge b$.
The \counterexample extraction procedures for these literals are defined similar to these of \chwellbehavedness.

\section{Extracting Liveness \Counterexamples}
\label{app:liveness}
In this section, we give details on how \tool handles liveness properties and their corresponding counterexamples. 

A liveness property is specified as a \buchi automaton \cite{buchi} or an LTL formula that accepts infinite traces in $\iglobalSemantics{\pcomplete}{c}$ violating the property.%\footnote{In fact, in \tool, the system designer can also specify liveness properties as (a restricted form of) Linear Time Logic (LTL) formulas that are then automatically translated to their \buchi automata equivalents.}
	
As is the case with safety properties, if the system is not live (i.e., $\iglobalSemantics{\pcomplete}{c} \not \models \livenessSpec$), a \counterexample $\cex \subseteq \iglobalSemantics{\pcomplete}{c}$ is extracted and encoded. Such $\cex$ is a ``lasso-shaped'' subset of the global semantics that represents an infinite, \fair, accepting trace violating $\livenessSpec$. An infinite trace is \fair if it represents a realistic infinite execution of the system as opposed to one that is merely an artifact of the non-determinism when picking which system event should be executed next. Such $\cex$ consists of two parts: a fair, accepting \emph{cycle} in $\iglobalSemantics{\pcomplete}{c}$, and a finite \emph{stem} from the initial state $\gq_0$ to a state in that cycle.
Note that $\cex$ \emph{additionally} contains all disabled global transitions from every state in the cycle. Intuitively, enabling any such disabled transition renders the infinite trace unfair, and hence rules out the liveness violation.

%In this section, 	Identifying fair, accepting cycles and their stems is a standard technique in the literature and we defer its details to \appref{liveness}.

We now discuss how liveness properties are checked in \tool. We first define our notion of \buchi automatons, the \emph{product structure} of a \buchi automaton and the global semantics $\globalSemantics{\process}{c}$, and the overall procedure of extracting witnesses of liveness violations.
 
 \subsection{\buchi Automatons and the Product Structure} 
 
 \para{\buchi Automatons.} A \buchi automaton is defined as $B= (S_B,s_{b0},T_B)$ where:
 \begin{compactenum}
 \item $S_B$ is the set of states which can optionally be accepting. $s_{b0}$ is the initial state.
  \item $T_B$  is the transition relation. Each transition $t_b \in T_B$ is of the form $(s_b, \event,\varphi_{q},s_b')$ where:
   \begin{compactenum}
     \item $s_b$ and $s_b'$ are the source and destination states, respectively,
     \item $\event \in \events$ is an event, and,
	\item $\varphi_{q}$ is a global state predicate.
 \end{compactenum}
  \end{compactenum}

 Note that this is slightly different than \buchi automatons with an event-only or state-only based labels on the transitions but one can think of any label on the edges as a “signal” the automaton needs to move from one state to the next.\ourskip

\para{Product Structure.} The product structure based on the global semantics  $\globalSemantics{\process}{c} = (Q , R)$ and \buchi automaton $B = (S_B,T_B)$ is defined as $PS = (S_{PS} \times T_{PS})$ where:
  \begin{compactenum}
\item $S_{PS} = Q \times S_B$ is the set of product states.
 \item $T_{PS} \subseteq S_{PS} \times S_{PS}$ is the transition relation. A transition $t_{PS} = (q,s_{B}) \rightarrow (q',s_{B}')$ is in $T_{PS}$ iff:
 \begin{compactenum}
      \item there exists a transition $(q,e,q')$ in $R$,
     \item there exists a transition $(s_B ,e, \varphi_q, s_B')$ in $T_B$, and,
     \item the state predicate holds in the destination state: $\varphi_q (q') = true$.
  \end{compactenum}
   \end{compactenum}

 \subsection{\Fair Infinite Traces}
 \para{Fairness.} We assume a notion of non-determinism that ensures the following:
\emph{if an event is ready infinitely often, it is taken infinitely often}. We say an event is \emph{ready} when processes are in local states where they can be acting/reacting to that event, and \emph{taken} if the event was chosen to be executed next.
 Intuitively, we want to ensure that infinite traces we report as liveness violations are not due to the fact that the nondeterministic choice of which system event to execute next is biased.

Then, infinite \fair accepting traces are identified using the following procedure:
\begin{compactenum}
\item Build the product structure $PS$.
\item Find fair accepting cycles (as detailed in \cite{Emerson.Utilizing.Symmetry.X.1997}).
\item Build a stem from the initial state to a state in the cycle.
\item extract the stem, the cycle, and all disabled transitions in every state on the cycle as a counter example to liveness.
\end{compactenum}

 \subsection{Specifying Liveness Properties in \aml}
To ease the designer's burden when specifying liveness properties, we extend \aml with two basic Liner-Time Logic (LTL) property templates that are automatically translated to \buchi automatons. The first property is $\mathbf{F} p$, requiring that proposition $p$ eventually holds in every execution in the system. The liveness property ``a leader is eventually elected'' is an example of this property. The second property is $\mathbf{G} (p \Rightarrow \mathbf{F} q)$, ensuring that every time proposition $p$ holds, then proposition $q$ eventually holds. The liveness property ``every command is eventually acknowledged'' is an example of this property.

Finally, we point out that we also check for \emph{deadlock-freedom}, i.e., that $\iglobalSemantics{\pcomplete}{c}$ does not contain any deadlock state where no event is enabled. A \counterexample to deadlock freedom is a trace to some deadlock state $\gq_d$ \emph{as well as} all disabled global transitions out $\gq_d$. After all, a deadlock state can be avoided if the system does not reach it, or can step out of it.

\section{Additional Evaluation Information}
\label{app:eval}

\subsection{Benchmarks Details}
In this section, for each benchmark, we provide a brief description of its functionality and its safety and liveness specifications. We also outline the unspecified functionality in each sketch.
%as well as the unspecified expressions in the initial sketch.
\begin{compactenum}
\item Distributed Store (DS) system: a distributed system where a set of processes maintain a consistent view of some stored data and manipulate the data according to client requests. A leader election protocol can be used to elect one process as a leader that is responsible for serving client requests while all other processes act as replicas.

\begin{compactenum}[ - ]
\item Safety properties: For all states in all behaviors, (i)
there is at most one leader, and (ii) the leader and the replicas agree on the stored data. 
\item Liveness properties: For all system behaviors, (i) a leader is eventually elected, and (ii) all client requests are eventually handled and acknowledged.

\item Unspecified functionality: the uninterpreted functions in the system's sketch correspond to various interesting questions including: (i)
how many candidates should win, and what should the winning and losing processes do after the partition?, (ii) how does a leader handle different types of client requests?, and (iii) how are the agreed-upon updates performed consistently across the leaders and replicas?
\end{compactenum}

\item Consortium (CTM): a distributed system where a set of processes try to make a decision by delegating decision making to an elected, trusted subset of deliberating processes. Once deliberating processes reach consensus on a decision, one of them is elected to broadcast the decision to the rest of the processes. 
\begin{compactenum}[ - ]
\item Safety properties: (i) at most two processes can participate in the deliberation stage and (ii) all the processes agree on the decision made. 
\item Liveness properties: (i) eventually, a set of deliberating processes is elected, and (ii) eventually, all processes in the system are informed about the decision made.
\item Unspecified functionality: the uninterpreted functions in the system's sketch correspond to various interesting questions including: (i) how should the processes partition themselves into deliberating processes and processes waiting for a decision to be made? (ii) once the decision is made, how should it be shared with the rest of the system? (iii) upon obtaining the decided values, how should the waiting processes behave? %\swen{maybe start newlines for enumerated questions?}
\end{compactenum}

\item Distributed Lock Service (DLS): a system that allows its clients to view a distributed file system as a centralized one through coarse-grain locking. The system elects a leader server that handles and acknowledges the clients requests, and ensures that the files remain consistent across all the replica servers. The leader can decide to step down, allowing other replica servers to take its place through a new election round.

\begin{compactenum}[ - ]
\item Safety property: there is at most one leader at any time.
\item Liveness properties: (i) a leader is eventually elected, and (ii) every client request is eventually served and acknowledged. 
\item Unspecified functionality: 
%\strikee{the process sketch of this system leaves some questions unanswered including:} 
how should the candidate servers elect a leader? how should they behave when the leader steps down?
\end{compactenum}

\item Distributed Register (DR): a distributed system that replicates a piece of data and allows concurrent updates to such data without an elected leader. Processes in the system can directly serve clients' read requests. If any process in the system receives an update request from its clients, and the value is different from the already-stored one, it invokes a round of consensus to update the data on all other processes. 
\begin{compactenum}[ - ]
\item Safety property: upon serving client read requests, all processes agree on the stored data.
\item Liveness property: all client requests are eventually acknowledged.
\item Unspecified functionality: 
%\strikee{the unspecified expressions in the sketch correspond to} 
the logic for deciding when an update request should be propagated to the rest of the processes, and how a process serves a read request.
\end{compactenum}

\item Two-Object Tracker (TOT): a surveillance system where a set of processes are constantly monitoring objects near them in the environment. If one object is observed, a leader process is elected to monitor the object along with some followers. If a second object is observed, another leader is elected to track that starts tracking that object and the followers are split between the two leaders. 

\begin{compactenum}[ - ]
\item Safety property: at most one leader monitoring each object.
\item Liveness property: if an object is observed by the system, then eventually a leader and some of the followers are tracking it.
\item Unspecified functionality: 
%\strikee{the unspecified expressions in the sketch correspond to} 
the logic of electing leaders and determining followers when objects are observed.
\end{compactenum}

\item Distributed Robot Flocking (DRF): a system where processes simulate moving in a flock by periodically electing a leader that decides the direction that the flock goes towards. 

\begin{compactenum}[ - ]
\item Safety property: at any given time, all processes are going in the same direction.
\item Liveness property: a leader is eventually elected.
\item Unspecified functionality: 
%\strikee{the unspecified expressions in the sketch correspond} 
the logic followers use to respond to a leader's change-of-direction messages.
\end{compactenum}

\item Small Aircraft Transportation System Landing Protocol (SATS): a protocol developed by NASA to enhance access to small airports that do not have a control tower. The aircraft coordinate with each other to access the airport and gradually descend to their final approach and landing. The airport's vicinity has two ``holding zones'' for aircraft to wait in before attempting to land. During landing, if the pilot observes any obstacles, they abort the landing and return to their appropriate holding zone.

\begin{compactenum}[ - ]
\item Safety properties: 
%\strikee{to avoid collisions, the protocol requires the following safety properties:}
 (i) at most four aircraft are allowed at a time in the airport's vicinity, (ii) there are at most two aircraft in each holding zone at a time, and (iii) at most one aircraft is allowed to attempt final approach at a time.
\item Liveness property: eventually some aircraft lands.  
\item Unspecified functionality: 
%\strikee{the unspecified expressions in the sketch correspond to} 
how the successive subsets of the aircraft are chosen to transition from entering the airport all the way until landing, and how aircraft behave if obstacles are noticed during landing.
\end{compactenum}

 We also study a variant of SATS, denoted SATS2, where the aircraft that misses its final approach is given a higher priority to attempt landing again.

\item Distributed Sensor Network (DSN): a system where a set of distributed sensors monitors the environment for signals (e.g., a smoke) and those that detect the signal coordinate to report it to a centralized authority.

\begin{compactenum}[ - ]
\item Safety property: at most two sensors send a report. 
\item Liveness property: if a signal is detected, then it is eventually reported. 
\item Unspecified functionality: 
%\strikee{the unspecified expressions in the sketch correspond to}
 various transitions the sensors must take to send a report. 
\end{compactenum}
We additionally study a variant of DSN, denoted DSNR, where the sensors can be reset back to their monitoring states.

\item Robotics Motion Planner (RMP): a system where a set of robots carries out tasks submitted by the environment. The robots need to coordinate to determine a motion plan that avoids collisions when executing their assigned tasks.

\begin{compactenum}[ - ]
\item Safety property: at most one robot can plan at a time, allowing it to account for the routes taken by previous robots. 
\item Liveness property: if a task is submitted, then it eventually is executed. 
\item Unspecified functionality: how the robots invoke consecutive rounds of agreement to pick a robot that can plan while others await. 
\end{compactenum}
Similar to DSN, we study a variant of RMP, denoted RMPR, where robots can be reset back to their initial states states by the environment.\ourskip

\end{compactenum}

\subsection{Complete Example}

We now illustrate \aml sketches and their completions using the Distributed Store benchmark.

\begin{figure*}[t]
\centering
\begin{tcolorbox}[colback=white,sharp corners,boxrule=0.3mm,top=-1mm,bottom=-1mm]
  %,top=-1.5mm,bottom=-1.5mm,right=0mm]
%\begin{tcolorbox}
\hspace{1em}\begin{minipage}{0.55\textwidth}
\begin{ssdsl}
machine DistributedStore
variables 
  int[1,5] cmd
  int[1,2] stored
actions
  env rz doCmd   : int[1,5]
  env rz ackCmd  : int[1,5]
  env rz ret  : int[1,2]
initial location Candidate (*\label{line:cand}*)
  on partition<elect>(All, (*\figHole{1}*)) // 1 (*\label{line:elect}*)
    win:  goto (*\figHole{2}*) // Leader
    lose: goto (*\figHole{3}*) // Replica (*\label{line:electe}*)
location Leader
  on recv(doCmd) do (*\label{line:request}*)
    cmd (*$\coloneqq$*) doCmd.payld
    if((*\figHole{4}*))  // cmd = 3
      goto (*\figHole{5}*) // Return
    else 
      goto (*\figHole{6}*) // RepCmd
location Return
  on _ do 
    rend(ret[stored], doCmd.sID)
    goto Leader
\end{ssdsl}
\end{minipage}
\begin{minipage}{0.70\textwidth}
 \lstset{firstnumber=25}
\begin{ssdsl}
location RepCmd
  on consensus<vcCmd>(All,1,cmd) do (*\label{line:cons1}*)
    cmd (*$\coloneqq$*) vcCmd.decVar[1] (*\label{line:dec1}*)
    if(cmd <= 2)  (*\label{line:ex1s}*)
      stored (*$\coloneqq$*) cmd
    else if((*\figHole{7}*))   // cmd = 4
      stored (*$\coloneqq$*) stored + 1
    else if((*\figHole{8}*))  // cmd = 5
      stored (*$\coloneqq$*) stored - 1 (*\label{line:ex1e}*)
   rend(ackCmd[cmd],doCmd.sID) (*\label{line:ack}*)
   cmd (*$\coloneqq$*) default(cmd)
   goto Leader
location Replica
  on consensus<vcCmd>(All,1,_) do (*\label{line:cons2}*)
    cmd (*$\coloneqq$*) vcCmd.decVar[1] (*\label{line:dec2}*)
    if(cmd <= 2) (*\label{line:ex2s}*)
      stored (*$\coloneqq$*) cmd
   else if(cmd = 4)  
     stored (*$\coloneqq$*) (*\figHole{9}*) // stored + 1
   else if(cmd = 5)
     stored (*$\coloneqq$*) (*\figHole{10}*) // stored - 1 (*\label{line:ex2e}*)
\end{ssdsl}
\end{minipage}
\end{tcolorbox}

%\begin{tcolorbox}[colback=white,sharp corners,boxrule=0.3mm, top=1mm, bottom = 2.5mm]
%\footnotesize{
% \textbf{Safety Property:} For all states in all system behaviors, there exists at most one leader.\\
% \textbf{Safety Property:} For all states in all system behaviors, the leader and the replicas agree on the stored data. \\
% \textbf{Liveness Property:} For all system behaviors, a leader is eventually elected.\\
%  \textbf{Liveness Property:} For all system behaviors, all client requests are eventually handled and acknowledged.}
%\end{tcolorbox}
\squeezecaption
\squeezecaption
\squeezecaption

\caption{\aml sketch of a Distributed Store process. %\neww{Assigning \termb{1},\termb{2},\termb{3},\termb{4}, or \termb{5}, respectively, to \termb{cmd} denote the commands: set data to 1, set data to 2, read the data, increment the data, or decrement the data, respectively.}\nour{or as in text?}
}
\label{fig:distStore} 
\squeezecaption
\squeezecaption
\squeezecaption
\squeezecaption
\squeezecaption
\squeezecaption

\end{figure*}

\para{Modeling Distributed Store in \aml}. 
The \aml model for Distributed Store is presented in \figref{distStore}. For now, assume that every symbol $\ufDSL{id}$ is replaced with the corresponding comment on the same line.

In the \aml model in \figref{distStore}, a process starts in the \termb{Candidate} location (\lineref{cand}) and coordinates with other processes to elect a leader.
%where they elect one process that acts as a leader, while %the rest of the processes act as replicas. 
Leader election is modeled using \aml's \terma{partition} agreement primitive (\lineref{elect}) that partitions a set of participants into ``winners'' and ``losers''. Here, \terma{partition} \termb{<elect>(All,1)} denotes a leader election round with identifier \termb{elect} where \termb{All} processes elect \termb{1} ``winning'' process that moves to the \termb{Leader} location, while all other ``losing'' processes move to the \termb{Replica} location. The leader awaits for client requests (\lineref{request}) and if it receives a read request (\termb{cmd = 3}), it moves to the \termb{Return} location and serves the request, and otherwise, it moves to the \termb{RepCmd} location to coordinate with the replicas on performing the requested data update. Agreement on the update to be performed on the data is modeled using \aml's \terma{consensus} primitive (Lines \ref{line:cons1} and \ref{line:cons2}) that allows a set of participants, each proposing one value, to agree on a set of decided values. Here, \terma{consensus}\termb{<vcCmd>(All,1,cmd)} denotes a consensus round with identifier \termb{vcCmd} where \termb{All} processes want to agree on \termb{1} decided value from the set of proposed values in the local variable \termb{cmd}. Upon termination of agreement, all processes retrieve the agreed-upon update to execute on the stored data using the \termb{vcCmd.}\terma{decVar}\termb{[1]} expression (Lines \ref{line:dec1} and \ref{line:dec2}), and execute the update on the stored data (Lines \ref{line:ex1s}-\ref{line:ex1e} and \ref{line:ex2s}-\ref{line:ex2e}).
The updates can be: set data to 1 (\termb{cmd = 1}), set data to 2 (\termb{cmd = 2}), increment the data (\termb{cmd = 4}), or decrement the data (\termb{cmd = 5}). 
%\nour{or as in caption?}
Once the update is performed on the stored data, the leader sends an acknowledgment back to the client (\lineref{ack}).\ourskip

\para{Distributed Store Sketch in \aml}. 
Consider the Distributed Store process sketch in \figref{distStore} where the system designer leaves parts of their model unspecified (denoted by the unspecified expressions $\ufDSL{id}$ for some \termb{id} in $\nats$). 
%\todo{motivate why these holes}
These unspecified parts correspond to various interesting questions the system designer may have when designing their system. For instance, unspecified expressions $\ufDSL{1}$, $\ufDSL{2}$, and $\ufDSL{3}$ determine
%describe the system in a way that makes the holes seem more interesting and involoved.: step 1 is partitioning, then we can say that synthesis does this. step 2 is leader logic on various commands? sythesis again. step 3: once we decide to replicate a command, both leader and replica do so in a way that maintains the cosistent veiew of the date. stthnesis gagin.
%\todo{Motivate the questions that the designer might have when thinking about such a system. This should be dome in a way that leads to the ``unspecified parts''. Currently,}
%determine the initial partition of candidates into two roles, and how many candidates should be in each role.
how many candidates should win, and what should the winning and losing processes do after the partition.
Unspecified expressions $\ufDSL{4}$, $\ufDSL{5}$, and $\ufDSL{6}$ dictate how a process acting as a leader handles different types of client requests. And finally,
unspecified expressions $\ufDSL{7}$, $\ufDSL{8}$, $\ufDSL{9}$, and $\ufDSL{10}$ control how the agreed-upon updates should be performed consistently across the leaders and replicas. 
%\todo{add the ``disclaimers'' not about control flow etc, but about dependability} 
While such unspecified expressions may not correspond to large segments of the code or complex control-flow constructs, they are the main ``knobs'' that the designer needs to turn to ensure that their systems belong to the \effdecidable fragment of \aml. The specified parts of this sketch capture a general intuition that an experienced system designer may have about this Distributed Store system. For instance, that there are leaders and replicas, and that different client requests may need to be handled differently.

\para{\tool}. From the set of 163,840,000 possible completions of the Distributed Store sketch in \figref{distStore}, \tool is able to find a correct one in 2.2 minutes and 203 iterations. In contrast, a naive exploration of all possible completions times out after 15 minutes. 
This demonstrates the impact of \counterexample extraction and encoding on effectively pruning the search space. During the execution of the synthesis loop, \tool encountered 36 \chwellbehavedness violations, 31 \amenability violations, 21 safety violations, and 114 liveness violations. Interestingly, out of the 67 \chwellbehavedness and \amenability violations, 63 were encountered in the first 100 iterations, showing that, for this system, \tool was able to converge faster on candidates in the \effdecidable fragment, and used liveness properties to pick a higher-quality candidate to return to the designer.

\section{Equivalent forms for the \CHwellbehavedness Conditions}
\label{app:prenexforms}
\begin{lemma}
\begin{multline*}
\forall \act,\actb \in \gsactions, s,s',t \in \LS: \\ 
\Big{(} \trans{s}{A(\act)}{s'} \wedge  \trans{s'}{R(\actb)} {*} \Big{)} 
 \Rightarrow
 \Big{(}
   \firableIn{\destSet{\act}}{\actb} \Rightarrow
   \Big{(}\trans{t}{R(\act)}{t'} \Rightarrow 
     \exists t' \xrightsquigarrow{R(\actb)} *
   \Big{)} 
 \Big{)}
 \end{multline*}
$\equiv$ 
\begin{multline*}
\forall \act,\actb \in \gsactions, s,s',t,z,z' \in \LS: \\ 
(\trans{s}{A(\act)}{s'} \wedge
\trans{s'}{R(\actb)} {*} \wedge
\trans{z}{A/R(\act)}{z'} \wedge 
\trans{z'}{A(\actb)}{*} \wedge 
\trans{t}{R(\act)}{t'}) 
\Rightarrow 
\exists t' \xrightsquigarrow{R(\actb)} *
\end{multline*}
\label{lem:prenexform2}
\end{lemma}
\begin{footnotesize}
\begin{proof}
We show this in one direction, the other can be obtained using the same steps in the reverse order. We start with:
\begin{multline*}
\forall \act,\actb \in \gsactions, s,s',t \in \LS:\\  
\Big{(} \trans{s}{A(\act)}{s'} \wedge  \trans{s'}{R(\actb)} {*} \Big{)} 
 \Rightarrow 
 \Big{(}
   \firableIn{\destSet{\act}}{\actb} \Rightarrow
   \Big{(}\trans{t}{R(\act)}{t'} \Rightarrow 
     \exists t' \xrightsquigarrow{R(\actb)} *
   \Big{)} 
 \Big{)}
\end{multline*}
Unrolling all the implications, we get:
\begin{multline*}
\forall \act,\actb \in \gsactions, s,s',t \in \LS:\\ 
\neg \Big{(} \trans{s}{A(\act)}{s'} \wedge  \trans{s'}{R(\actb)} {*} \Big{)} 
 \vee
 \Big{(}
   \neg \firableIn{\destSet{\act}}{\actb} \vee
  \neg \Big{(}\trans{t}{R(\act)}{t'} \vee 
     \exists t' \xrightsquigarrow{R(\actb)} *
   \Big{)} 
 \Big{)}
 \end{multline*}
 \begin{multline*}
= \forall \act,\actb \in \gsactions, s,s',t \in \LS: \\ 
\neg \trans{s}{A(\act)}{s'} \vee 
\neg \trans{s'}{R(\actb)} {*} \vee
\neg \firableIn{\destSet{\act}}{\actb} \vee
\neg \trans{t}{R(\act)}{t'} \vee 
\exists t' \xrightsquigarrow{R(\actb)} *
\end{multline*}
\begin{multline*}
 = \forall \act,\actb \in \gsactions, s,s',t \in \LS: \\
 \neg 
 \Big{(}
 \trans{s}{A(\act)}{s'} \wedge
  \trans{s'}{R(\actb)} {*} \wedge
  \firableIn{\destSet{\act}}{\actb} \wedge
  \trans{t}{R(\act)}{t'}  
 \Big{)} \vee 
 \exists t' \xrightsquigarrow{R(\actb)} *
\end{multline*}
 \begin{multline*}
 = \forall \act,\actb \in \gsactions, s,s',t \in \LS: \\
 \Big{(}
 \trans{s}{A(\act)}{s'} \wedge
  \trans{s'}{R(\actb)} {*} \wedge
  \firableIn{\destSet{\act}}{\actb} \wedge
  \trans{t}{R(\act)}{t'}  
 \Big{)} \Rightarrow
 \exists t' \xrightsquigarrow{R(\actb)} *
\end{multline*}
Recall that
$$\firableIn{\destSet{\act}}{\actb} \equiv \exists z,z' \in \LS:  \trans{z}{A/R(\act)}{z'} \wedge 
 \trans{z'}{A(\actb)}{*}$$
Hence our formula becomes 
 \begin{multline*}
 = \forall \act,\actb \in \gsactions, s,s',t \in \LS:  \\
 \Big{(}
 \trans{s}{A(\act)}{s'} \wedge
  \trans{s'}{R(\actb)} {*} \wedge
(\exists z,z' \in \LS:  \trans{z}{A/R(\act)}{z'} \wedge 
 \trans{z'}{A(\actb)}{*}) \wedge
  \trans{t}{R(\act)}{t'}  
 \Big{)} \Rightarrow
 \exists t' \xrightsquigarrow{R(\actb)} *
  \end{multline*}
Since all the variables in $\Big{(}
 \trans{s}{A(\act)}{s'} \wedge
  \trans{s'}{R(\actb)} {*} \wedge
  \trans{t}{R(\act)}{t'}  
 \Big{)}$ are free\footnote{Modulo some renaming for event names if need be.} in $(\exists z,z' \in \LS:  \trans{z}{A/R(\act)}{z'} \wedge 
  \trans{z'}{A(\actb)}{*})$, we can apply the following rewrite rule: $(\exists x:f(x)) \wedge g(y)  \equiv \exists x: (f(x) \wedge g(y))$: 
   \begin{multline*}
   \forall \act,\actb \in \gsactions, s,s',t \in \LS:  \\
\Big{(}
     \exists z,z' \in \LS: ( \trans{s}{A(\act)}{s'} \wedge
    \trans{s'}{R(\actb)} {*} \wedge
    \trans{z}{A/R(\act)}{z'} \wedge 
   \trans{z'}{A(\actb)}{*} \wedge
    \trans{t}{R(\act)}{t'})  
   \Big{)} \Rightarrow
   \exists t' \xrightsquigarrow{R(\actb)} *
    \end{multline*}
Following a similar variable freeness argument as before, we can apply the following rewrite rule: $(\exists x:f(x)) \Rightarrow g(y)  \equiv \forall x: (f(x) \Rightarrow g(y))$: 
   \begin{multline*}
   \forall \act,\actb \in \gsactions, s,s',t \in \LS:  \\
     \forall z,z' \in \LS: \Big{(} (\trans{s}{A(\act)}{s'} \wedge
    \trans{s'}{R(\actb)} {*} \wedge
    \trans{z}{A/R(\act)}{z'} \wedge 
   \trans{z'}{A(\actb)}{*} \wedge
    \trans{t}{R(\act)}{t'})  
 \Rightarrow
   \exists t' \xrightsquigarrow{R(\actb)} *    \Big{)}
    \end{multline*}
Finally, we get:
\begin{multline*}
\forall \act,\actb \in \gsactions, s,s',t,z,z' \in \LS:  \\
(\trans{s}{A(\act)}{s'} \wedge
\trans{s'}{R(\actb)} {*} \wedge
\trans{z}{A/R(\act)}{z'} \wedge 
\trans{z'}{A(\actb)}{*} \wedge 
\trans{t}{R(\act)}{t'}) 
\Rightarrow 
\exists t' \xrightsquigarrow{R(\actb)} *
\end{multline*}
    
\end{proof}
\end{footnotesize}

\begin{lemma}
\begin{multline*}
\forall \actb \in \gsactions, s,s' \in \LS: 
\Big{(} \trans{s}{}{s'} \wedge \trans{s'}{R(\actb)}{*} \Big{)} \Rightarrow \\
\Big{(}
\forall X \in \pha(s): 
\Big{(}
\firableIn{X}{\actb} \Rightarrow  
\big{(} \forall t \in X.\ \exists t \xrightsquigarrow{R(\actb)} * \big{)}
\Big{)}
\Big{)}
\end{multline*}
$\equiv$
\begin{multline*}
= \forall \actb \in \gsactions, s,s',t,z \in \LS, X \in \pha(s): \\ 
\big{(}
\samePhase{X}{s}{t} \wedge 
\samePhase{X}{s}{z} \wedge
\trans{s}{}{s'} \wedge 
\trans{s'}{R(\actb)}{*} \wedge
\trans{z}{A(\actb)}{*} 
\big{)}
\Rightarrow
\big{(}\exists t \xrightsquigarrow{R(\actb)} * \big{)}
\end{multline*}
\label{lem:prenexform3}
\end{lemma}

\begin{proof}
\begin{footnotesize}
We show this in one direction, the other can be obtained using the same steps in the reverse order. We start with:
\begin{multline*}
\forall \actb \in \gsactions, s,s' \in \LS: \\
\Big{(} \trans{s}{}{s'} \wedge \trans{s'}{R(\actb)}{*} \Big{)} \Rightarrow
\Big{(}
\forall X \in \pha(s): 
\Big{(}
\firableIn{X}{\actb} \Rightarrow  
\big{(} \forall t \in X.\ \exists t \xrightsquigarrow{R(\actb)} * \big{)}
\Big{)}
\Big{)}
\end{multline*}
Unrolling the implications, we get:
\begin{multline*}
\forall \actb \in \gsactions, s,s' \in \LS: \\
\neg \Big{(} \trans{s}{}{s'} \wedge \trans{s'}{R(\actb)}{*} \Big{)} 
\vee
\Big{(}
\forall X \in \pha(s): 
\Big{(}
\neg \firableIn{X}{\actb} 
\vee
\big{(} \forall t \in X.\ \exists t \xrightsquigarrow{R(\actb)} * \big{)}
\Big{)}
\Big{)}
\end{multline*}	
\begin{multline*}
= \forall \actb \in \gsactions, s,s' \in \LS: \\
\neg \trans{s}{}{s'} \vee \neg \trans{s'}{R(\actb)}{*} 
\vee
\Big{(}
\forall X \in \pha(s): 
\Big{(}
\neg \firableIn{X}{\actb} 
\vee
\big{(} \forall t \in X.\ \exists t \xrightsquigarrow{R(\actb)} * \big{)}
\Big{)}
\Big{)}
\end{multline*}
Using the rewrite rule: $(\forall x:f(x)) \vee g(y) \equiv \forall x:(f(x) \vee g(y))$ we obtain the following formula:
\begin{multline*}
\forall \actb \in \gsactions, s,s' \in \LS: \\
\forall X \in \pha(s): 
\Big{(}
\neg \trans{s}{}{s'} \vee \neg \trans{s'}{R(\actb)}{*} 
\vee
\neg \firableIn{X}{\actb} 
\vee
\big{(} \forall t \in X.\ \exists t \xrightsquigarrow{R(\actb)} * \big{)}
\Big{)}
\end{multline*}
Using the same rewrite rule we get:
\begin{multline*}
\forall \actb \in \gsactions, s,s' \in \LS: \\
\forall X \in \pha(s), t \in X: 
\Big{(}
\neg \trans{s}{}{s'} \vee \neg \trans{s'}{R(\actb)}{*} 
\vee
\neg \firableIn{X}{\actb} 
\vee
\big{(}\exists t \xrightsquigarrow{R(\actb)} * \big{)}
\Big{)}
\end{multline*}
Then
\begin{multline*}
\forall \actb \in \gsactions, s,s' \in \LS, X \in \pha(s), t \in X: \\ 
\Big{(}
\neg \trans{s}{}{s'} \vee \neg \trans{s'}{R(\actb)}{*} 
\vee
\neg \firableIn{X}{\actb} 
\vee
\big{(}\exists t \xrightsquigarrow{R(\actb)} * \big{)}
\Big{)}
\end{multline*}
Recall that $\samePhase{P}{u}{v}$ indicates that states $u$ and $v$ are in some phase $P$ together. Hence, we can rewrite $t \in X$ as follows:
\begin{multline*}
\forall \actb \in \gsactions, s,s',t \in \LS, X \in \pha(s): \\ 
\samePhase{X}{s}{t} \Rightarrow \Big{(}
\neg \trans{s}{}{s'} \vee \neg \trans{s'}{R(\actb)}{*} 
\vee
\neg \firableIn{X}{\actb} 
\vee
\big{(}\exists t \xrightsquigarrow{R(\actb)} * \big{)}
\Big{)}
\end{multline*}
\begin{multline*}
= \forall \actb \in \gsactions, s,s',t \in \LS, X \in \pha(s): \\ 
\Big{(} \neg \samePhase{X}{s}{t} \vee 
\neg \trans{s}{}{s'} \vee \neg \trans{s'}{R(\actb)}{*} 
\vee
\neg \firableIn{X}{\actb} 
\vee
\big{(}\exists t \xrightsquigarrow{R(\actb)} * \big{)}
\Big{)}
\end{multline*}
\begin{multline*}
= \forall \actb \in \gsactions, s,s',t \in \LS, X \in \pha(s): \\ 
\Big{(} \neg \big{(}\samePhase{X}{s}{t} \wedge 
 \trans{s}{}{s'} 
 \wedge \trans{s'}{R(\actb)}{*} 
\wedge
\firableIn{X}{\actb} \big{)}
\vee
\big{(}\exists t \xrightsquigarrow{R(\actb)} * \big{)}
\Big{)}
\end{multline*}
\begin{multline*}
= \forall \actb \in \gsactions, s,s',t \in \LS, X \in \pha(s): \\ 
\Big{(} \big{(}\samePhase{X}{s}{t} \wedge 
 \trans{s}{}{s'} 
 \wedge \trans{s'}{R(\actb)}{*} 
\wedge
\firableIn{X}{\actb} \big{)}
\Rightarrow
\big{(}\exists t \xrightsquigarrow{R(\actb)} * \big{)}
\Big{)}
\end{multline*}
Finally, we can rewrite $\firableIn{X}{\actb}$ using the following:
$$\firableIn{X}{\actb} \equiv \exists z \in X : \trans{z}{A(\actb)}{*} \equiv \exists z \in S : \samePhase{X}{s}{z} \wedge \trans{z}{A(\actb)}{*} $$
Hence, we obtain: 
\begin{multline*}
= \forall \actb \in \gsactions, s,s',t \in \LS, X \in \pha(s): \\ 
\Big{(} \big{(}\samePhase{X}{s}{t} \wedge 
 \trans{s}{}{s'} 
 \wedge \trans{s'}{R(\actb)}{*} \\
\wedge
(\exists z \in S : \samePhase{X}{s}{z} \wedge \trans{z}{A(\actb)}{*}) \big{)}
\Rightarrow
\big{(}\exists t \xrightsquigarrow{R(\actb)} * \big{)}
\Big{)}
\end{multline*}
Following similar steps from \lemref{prenexform2} we extract $z$ to the top level and obtain:
\begin{multline*}
= \forall \actb \in \gsactions, s,s',t,z \in \LS, X \in \pha(s): \\ 
\big{(}
\samePhase{X}{s}{t} \wedge 
\samePhase{X}{s}{z} \wedge
\trans{s}{}{s'} \wedge 
\trans{s'}{R(\actb)}{*} \wedge
\trans{z}{A(\actb)}{*} 
\big{)}
\Rightarrow
\big{(}\exists t \xrightsquigarrow{R(\actb)} * \big{)}
\end{multline*}

\end{footnotesize}

\end{proof}

\end{document}